\documentclass[preprint,3p,twocolumn]{elsarticle}

\usepackage{hyperref,lineno}
\modulolinenumbers[5]

\journal{Nuclear Instruments and Methods in Physics Research A}

\newcommand{\gm}{$g$\,$-$\,$2$}
\newcommand{\pb}{PbF$_2$}
\newcommand{\mtca}{$\mu$TCA}

\begin{document}
\begin{frontmatter}
\title{Performance of the Muon $g-2$ calorimeter and readout systems \\measured with test beam data}

\author[uw]{K.\,S.~Khaw\corref{ksknote}}
\cortext[ksknote]{Corresponding author: Tel.: +1-206-543-2996.}
\ead{khaw84@uw.edu}

\author[genova]{M.~Bartolini}
\author[uw]{H.\,~Binney}
\author[cornell]{R.\,Bjorkquist}

\author[cornell]{A.\,Chapelain}

\author[uudine,trieste]{A.~Driutti}

\author[pisa1,pisa2]{C.~Ferrari}
\author[uw]{A.\,T.~Fienberg}
\author[pisa1,pisa2]{A.~Fioretti}

\author[pisa1,pisa2]{C.~Gabbanini}
\author[uiuc]{S.~Ganguly}
\author[cornell]{L.\,K.~Gibbons}
\author[roma]{A.~Gioiosa}
\author[jmu]{K.~Giovanetti}
\author[uky]{W.\,P.~Gohn}
\author[uky]{T.\,P.~Gorringe}

\author[uw]{J.\,B.~Hempstead}
\author[uw]{D.\,W.~Hertzog}

\author[napoli1,napoli2]{M.~Iacovacci}

\author[uw]{J.~Kaspar}
\author[uiuc]{A.~Kuchibhotla}

\author[uiuc]{S.~Leo}
\author[pisa2,pisa3]{A.~Lusiani}

\author[napoli2]{S.~Mastroianni}

\author[uudine,trieste]{G.~Pauletta}
\author[uw]{D.\,A.~Peterson}
\author[uva]{D.~Po\v{c}ani\'{c}}

\author[cornell]{N.~Rider}

\author[uiuc]{C.\,D.~Schlesier}
\author[uw,pisa2]{M.\,W.~Smith}
\author[ucl]{T.~Stuttard}
\author[cornell]{D.\,A.~Sweigart}

\author[uw]{T.\,D.~Van\,Wechel}
\author[pisa2]{G.~Venanzoni}

\address[uw]{University of Washington, Box 351560, Seattle, WA 98195, USA}
\address[genova]{Universit\`a di Genova and INFN, Sezione di Genova, Genova, Italy}
\address[cornell]{Cornell University, Ithaca, NY 14850, USA}
\address[uudine]{Universit\`a di Udine, Udine, Italy}
\address[trieste]{INFN Sezione di Trieste, Trieste, Italy}
\address[pisa1]{Istituto Nazionale di Ottica del C.N.R., UOS Pisa, Pisa, Italy}
\address[pisa2]{INFN, Sezione di Pisa, Pisa, Italy}
\address[uiuc]{University of Illinois at Urbana-Champaign, Urbana, IL 61801, USA}
\address[roma]{Universit\`a del Molise and INFN, Sezione di Roma Tor Vergata, Roma, Italy}
\address[jmu]{James Madison University, Harrisonburg, VA 22807, USA}
\address[uky]{University of Kentucky, Lexington, KY 40506, USA}
\address[napoli1]{Universit\`a di Napoli, Napoli, Italy}
\address[napoli2]{INFN, Sezione di Napoli, Napoli, Italy}
\address[pisa3]{Scuola Normale Superiore, Pisa, Italy}
\address[uva]{University of Virginia, Charlottesville, VA 22904, USA}
\address[ucl]{University College London, London WC1E 6BT, UK}

\begin{abstract}
A single calorimeter station for the Muon \gm\ experiment at Fermilab includes the following subsystems: a 54-element array of \pb\ Cherenkov crystals read out by large-area SiPMs, bias and slow-control electronics, a suite of 800\,MSPS waveform digitizers, a clock and control distribution network, a gain calibration and monitoring system, and a GPU-based front-end which is read out through a MIDAS data acquisition environment.
The entire system performance was evaluated using $2.5 - 5$\,GeV electrons at the End Station Test Beam at SLAC.
This paper includes a description of the individual subsystems and the results of measurements of the energy response and resolution, energy-scale stability, timing resolution, and spatial uniformity. All measured performances meet or exceed the \gm\ experimental requirements.
Based on the success of the tests, the complete production of the required 24 calorimeter stations has been made and installation into the main experiment is complete.
Furthermore, the calorimeter response measurements reported here informed the design of the reconstruction algorithms that are now employed in the running \gm\ experiment.

\end{abstract}

\begin{keyword}
Lead-fluoride calorimeter, Silicon photomultiplier, Waveform digitizer, Laser calibration
\PACS 29.40.V \sep 13.35.B \sep 14.60.E
\end{keyword}

\end{frontmatter} 


\section{Introduction}
\label{sec:intro}

The Muon \gm\ experiment E989~\cite{Grange:2015fou} at Fermi National Accelerator Laboratory (Fermilab) aims to determine the anomalous magnetic moment $a_{\mu} \equiv (g-2)/2$ of the muon to a relative precision of 140 parts per billion (ppb).
The measurement is made by observing the spin precession frequency $\omega_s$ relative to the cyclotron frequency $\omega_c$ for muons orbiting a highly uniform magnetic storage ring with field $\vec{B}$.
Expressed using above-mentioned quantities,
\begin{equation}
a_{\mu} = -(m_\mu / q)(\omega_a/B),
\label{eq:simpleomega}
\end{equation}
where the anomalous precession frequency is defined as $\omega_a \equiv \omega_s - \omega_c$.
The experiment aims to measure $\omega_a$ to a statistical precision of 100\,ppb, with $\omega_a$ and $B$ systematics determined to 70\,ppb each.

This paper describes the final prototype instrumentation developed to measure $\omega_a$.
It is organized as follows.
Section \ref{sec:gm2requirements} outlines how the measurement of the anomalous precession frequency is carried out and the consequent technical demands on the instrumentation that guided the design decisions.
Section \ref{sec:instrumentation} describes each of the key calorimeter subsystems.
Section \ref{sec:setup} describes the experimental setup at SLAC test beam. Section \ref{sec:recon} outlines the reconstruction and calibration methods.
Section \ref{sec:performance} provides system performance metrics such as timing resolution, linearity, energy resolution, and spatial uniformity.
The measurements and analysis described herein are based on a three-week run using the End Station Test Beam (ESTB) at SLAC in June 2016.
The final production of the complete calorimeter system is now installed and running in the Muon \gm\ experiment at Fermilab.


\section{Measurement of the Anomalous Precession Frequency}
\label{sec:gm2requirements}

The design of the Muon \gm\ experiment at Fermilab follows largely the well-known method employed most recently in the E821 experiment~\cite{Bennett:2006fi} at Brookhaven National Laboratory (BNL).
Intense bunches of polarized positive muons having a central momentum of 3.1\,GeV/c are injected into the BNL 7.11\,m radius, 1.45\,T superconducting storage ring~\cite{Danby:2001eh} that was relocated to Fermilab. 
The beam has negligible hadron contamination, but non-negligible fraction of positrons.  
Once injected, the particles receive a magnetic kick to deflect them onto a stable orbit. 
Only a few percent of the muons remain after a few turns (cyclotron period is 149.2\,ns).
The unstored muons, together with the positron beam contamination, generate a prompt background in the calorimeter systems that is anticipated in the detector and electronics designs.
The storage ring fill cycle is repeated at an average rate of 11.4\,Hz.

Stored muons orbit the ring and their spins precess according to Eq.~\ref{eq:simpleomega}.
The $\omega_a$ frequency, which is the subject of the instrumentation described here, is encoded naturally in the time and energy distribution of decay positrons.
Because parity violation in $\mu^+ \rightarrow e^+{\bar{\nu}}_{\mu}\nu_e$ associates the decay positron energy in the laboratory frame to the average muon spin direction at the time of decay, the higher-energy positrons are preferentially emitted when the muon spin is aligned with its momentum, and lower-energy positrons are emitted when the spin is reversed.
Therefore the number $N$ of higher-energy positrons striking detectors follows a functional form
\begin{equation}
N(t) = N_{0}\exp(-t/\gamma\tau)[1 + A \cos (\omega_a t + \phi)]
\end{equation}
where $N_{0}$ is a normalization, $\gamma\tau$ is the time-dilated muon lifetime ($\sim64.4\,\mu$s), $A$ is the decay asymmetry, and $\phi$ is an arbitrary phase.

\begin{figure*}[htbp]
\centering
\includegraphics[width=\linewidth]{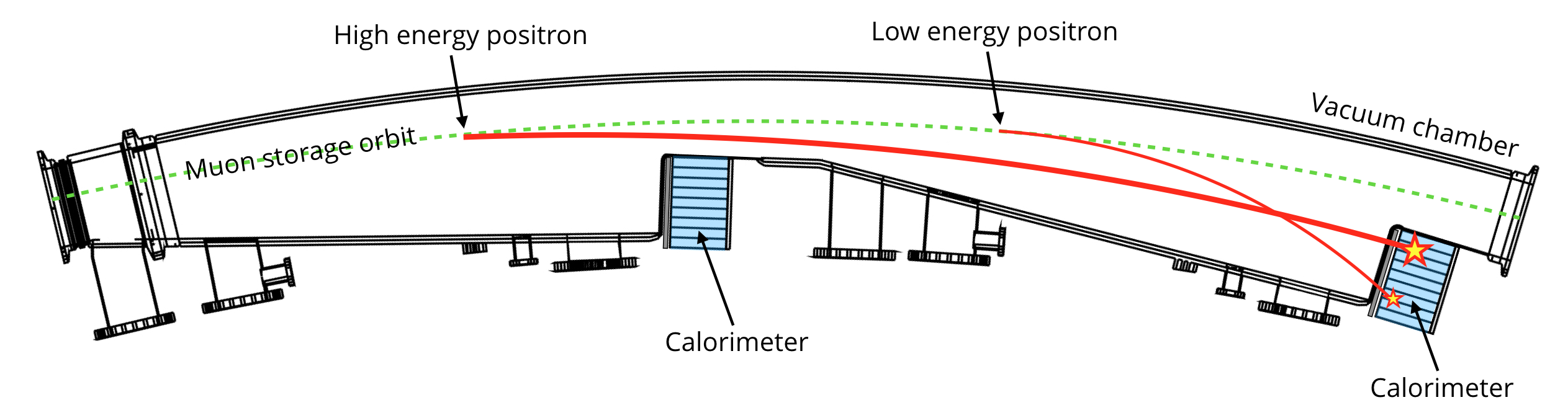}
\caption{\label{fig:calonearVC} Scalloped vacuum chamber with positions
of calorimeter stations indicated. A high-(low-) energy decay positron trajectory is shown by the thick (thin) line, which impinges on the front face of the calorimeter array.}
\end{figure*}

In practice, 24 calorimeter stations are positioned evenly around the inner radius of the storage ring, each being tucked into a notch of the scalloped vacuum chamber system. 
Figure~\ref{fig:calonearVC} shows the locations of two calorimeters with respect to a segment of the storage ring.
The stored muons are constrained to occupy a 9-cm diameter cross-sectional area within the vacuum chamber. 
The decay positrons have momenta below the muon momenta and therefore curl to the inside of the ring. 
They exit the vacuum chamber through a thin aluminum wall directly adjacent and parallel to the front face of the calorimeter.

\subsection{Technical Requirements of the Calorimeter System}

We define the Calorimeter System to include the following subsystems: the physical calorimeter, its readout transducers, its electronics and controls infrastructure, the waveform digitizers, a clock and control distribution network, a laser gain calibration and monitoring system, and a GPU-based frontend readout coupled to a data acquisition environment.
The system must record data in a dead-time free manner throughout the $\sim700\,\mu$s muon fills after receiving an accelerator trigger. 
Each calorimeter station produces $\sim800$\,MByte/s of raw digitized samples, which must be filtered using an online GPU farm so that only regions, dubbed ``islands,'' that have at least an over-threshold hit are saved for offline analysis.

The leading $\omega_a$ systematic uncertainties associated with the calorimeter are multi-particle pileup and gain stability. 
Both are exacerbated by the instantaneous rate that can exceed 10\,MHz just after injection, but then drops by a factor of $\sim50,000$ over the course of a fill. 
Because of muon beam cleaning and debunching requirements, the typical start-time of a fit for $\omega_a$ is $30\,\mu$s after injection. 
A combination of lessons learned from the BNL experiment and various Geant4~\cite{Agostinelli:2002hh} simulations establishes several performance parameters that must be realized. 
These, along with practical considerations, motivate the following list of system requirements:
\begin{itemize}
\item The energy resolution at 2\,GeV should be better than 5\,\%.
\item The calorimeter gain must recover by $30\,\mu$s following the intense flash at injection.
\item The calorimeter gain must remain stable during the $30 - 700\,\mu$s measurement period.
\item The laser calibration system must be able to correct for residual gain instabilities to better than $4 \times 10^{-4}$ during the measurement period.
\item The time resolution of a reconstructed shower should be better than 100\,ps for positrons with energy greater than 1.8\,GeV.
\item The time stability of the system must be better than 7\,ps throughout any fill, to ensure less than a 10\,ppb shift to $\omega_a$.
\item The calorimeter must be able to resolve two electromagnetic showers with impact time separations greater than 5\,ns with 100\% efficiency.
\item The calorimeter must fulfill the geometrical space limitation by fitting into the scallop of the vacuum chamber.
\item The calorimeter, readout, and cabling that reside adjacent to a highly uniform magnetic field must function in the field and, importantly, must not perturb the field uniformity.
\end{itemize}


\section{Descriptions of the Calorimeter Subsystems}
\label{sec:instrumentation}

\subsection{Lead Fluoride Calorimeter}

\begin{figure*}[htbp]
\centering
\includegraphics[width=.405\linewidth]{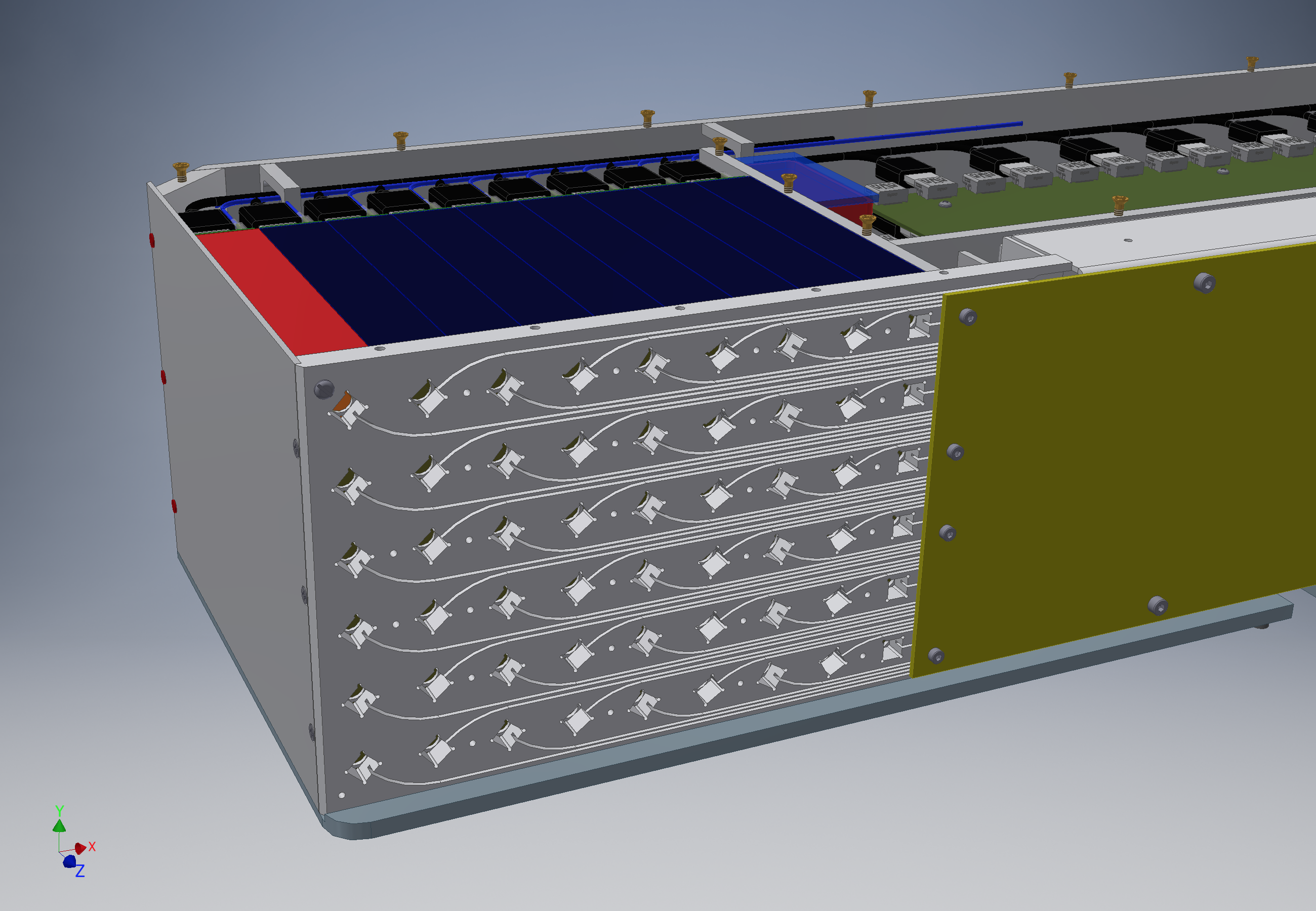}
\includegraphics[width=.585\linewidth]{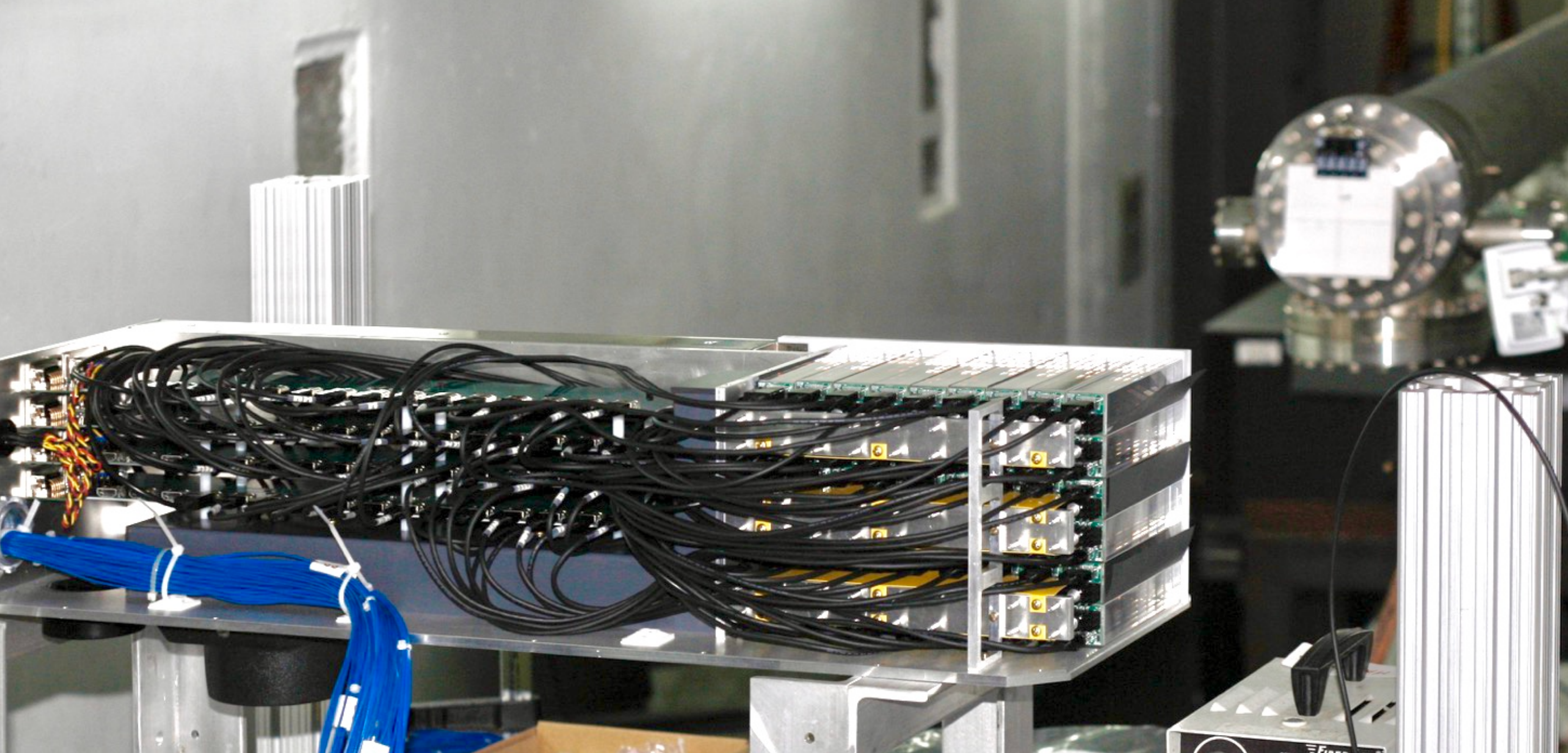}
\caption{\label{fig:gm2calo} Left: A CAD drawing of the calorimeter showing the exposed light distribution panel, the crystal array immediately behind it, the breakout boards with several HDMI connectors, and the support platform. Right: Photo of the back side of an open calorimeter that is sitting on top of an $x-y$ movable table. The electron beam exits the vacuum tube at the top-right corner of the figure and impinges on the front face of the calorimeter crystal array.}
\end{figure*}

An individual calorimeter station consists of fifty four $25 \times 25 \times 140$~mm$^{3}$ SICCAS\footnote{Shanghai SICCAS High Technology Corporation, 1295 Dingxi Rd., Shanghai 200050, China} \pb\ Cherenkov crystals stacked in a 9 wide by 6 high array, as shown in Fig.~\ref{fig:gm2calo} (left). 
Each crystal is wrapped in a single layer of matte black, non-reflective Tedlar\texttrademark on the four long sides.  
A $12 \times 12$ mm$^{2}$ Hamamatsu MPPC\footnote{Multi-Pixel Photon Counter Model number S12642-4040PA-50} (SiPM) is glued to the downstream face using an index-matching optical epoxy\footnote{Zeiss Resin OK 2030 and Hardener H 950}.
In Ref.~\cite{Fienberg:2014kka}, we documented performance tests associated with the choice of \pb\ Cherenkov crystals as a moderator for the calorimeter, and we explored two crystal wrapping alternatives. 
The use of Tedlar\texttrademark\,results in a lower light yield, but a narrower pulse shape, compared to a reflective white wrapping.
In Ref.~\cite{Kaspar:2016ofv}, we detailed the development of the SiPM electronics and the bias supply selection considerations, which had to be optimized for the high-rate and short-pulse-shape demands of this experiment.

The bias supply of the SiPMs is provided by four BK Precision 9124 commercial programmable DC power supplies.
Typical values of applied bias voltages $V_{\rm{ap}}$ are in the range of $66 - 68$\,V, being adjusted to set the overvoltage\footnote{Overvoltage, $V_{\rm{ov}}=V_{\rm{ap}}-V_{\rm{bd}}$, where $V_{\rm{ap}}$ is the applied voltage and $V_{\rm{bd}}$ is the breakdown voltage.} $V_{\rm{ov}}$ in the range of approximately 1\,V above the breakdown voltage $V_{\rm{bd}}$.
The choice of bias voltage values optimized the SiPM photo-detection efficiency (PDE), minimized the dark count rate, and allowed the gain to match the dynamic range of the electronics such that two 3~GeV coincident electrons would not saturate the digitizer dynamic range.

The low voltage for the SiPM pre-amplifier circuit boards is provided by an OTE HY3003-3 DC power supply.
The SiPM pre-amplifier boards are controlled using HDMI cables that connect each to a breakout board mounted within the box service compartment.
The breakout board distributes the bias voltage levels and the communication information between SiPM boards and a BeagleBone computer.
The housing around the crystal SiPM ends is cooled from the bottom by air fans and a duct-work corridor internal to the box.
Signals from the SiPMs are connected to the waveform digitizers through custom Samtec ECDP cables.
The positioning of the calorimeter relative to the ESTB beamline is shown in Fig.~\ref{fig:gm2calo} (right).

\subsection{Laser calibration and monitoring system}
\label{subsec:lasersystem}

\begin{figure}[htbp]
\centering
\includegraphics[width=\linewidth]{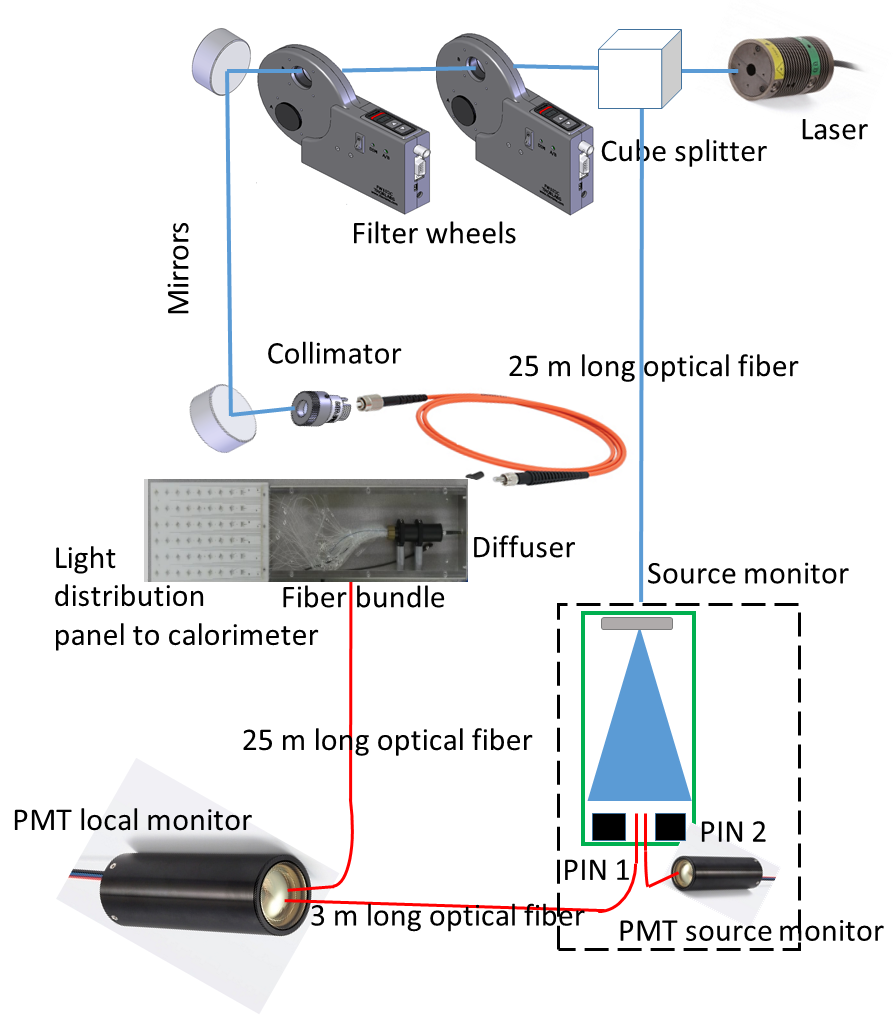}
\caption{\label{fig:LMSM} The laser system in detail: Laser head with 2 filter wheels for intensity adjustment, source monitor, optical fibers for sending the calibration pulses to the calorimeter and back to local monitor.}
\end{figure}

A SiPM is a temperature-sensitive device, and lab tests show that its breakdown voltage changes with a temperature coefficient of about 70\,mV/$^\circ$C. 
We found that at an operational over-voltage of 2.4\,V, the SiPM gain changes by 2.5\% per $^\circ$C; note that this overvoltage is higher than we nominally use.
To initially equalize the gains of the SiPMs and to monitor their drifts, a laser calibration system has been developed.
It is described in some detail in Ref.~\cite{Anastasi:2016luh} and its main features are briefly recalled here. 
This laser system also represents an important tool for debugging functionality of the calorimeters, their electronics, and the data acquisition system prior to the experimental run.
The laser calibration system deployed at this test beam is shown in Fig.~\ref{fig:LMSM}.

In the laser box, the light from a pulsed diode laser (405\,nm, PicoQuant LDH-P-C-405M, pulse length $<1$\,ns) is divided using an optical beam splitter with the first path directing 80\% of the intensity to the calorimeter station through a 25\,m long silica optical fiber. 
Upon arrival, it is uniformly spread and coupled into a bundle of plastic optical fibers by means of a diffuser.
These light pulses are conveyed into the upstream faces of the 54 \pb\ crystals using prisms embedded in a light distribution panel made of Delrin that forms the upstream wall of the calorimeter enclosure, see Fig.~\ref{fig:gm2calo} (left). 
A sample of this diffused light is returned to the laser box where it is measured by a local monitor (LM). 
The second light path directs the 20\% remaining light fraction to a source monitor (SM) and to the LM.

The SM is designed to measure the laser power directly after the laser head, minimizing the sensitivity to any perturbation occurring in all optical elements in the line.
It utilizes a sizable fraction of the laser light in order to reduce shot noise and reach the required statistical precision rapidly.
A specialized electronics module (Monitoring Board)~\cite{Anastasi:2018} was utilized to manage photo-detectors (two pin diodes - PIN1 and PIN2, and a PMT), signal processing and data readout for the three SM channels.
It amplifies and filters the signals, digitizes them and stores the data in local FIFO buffers. 
The back-end section collects the data frames from the buffer channels, performs the event-building and provides them to a PC server for data storage and further processing.
The three filtered signals, made fully Gaussian with 600\,ns width, are distributed to custom-designed waveform digitizers.

The LM, on the other hand, is designed to monitor the stability of the light distribution chain from the laser head to the optical fiber bundle after the diffuser. 
In particular, it is sensitive to any variance in the coupling and the transmission of the laser into the silica fiber and to the splitting inside the diffuser.
The fast PMT in the LM sees two comparable pulses separated by 250\,ns. The first pulse is coming from the SM; the second corresponds to light having traveled approximately 50\,m from the laser to the calorimeter and back.
In this way, the LM allows the monitoring and the correction of instabilities introduced by most of the light distribution components.


\subsection{Clock and Calorimeter Back-end Electronics}

The signal processing of waveforms produced by the calorimeter must provide high fidelity determination of both the time-of-arrival and the energy of the decay positron.
The electronics must also be as robust as possible against potential rate-dependent biases in positron selection.
To meet these needs, we have produced a five-channel custom-designed waveform digitizer (WFD5)~\cite{Sweigart:2016jty}.
The WFD5 conforms to the Advanced Mezzanine Card (AMC) $\mu$TCA\textsuperscript{\textregistered} standard and functions with the AMC13 module~\cite{Hazen:2013rma} designed by Boston University (BU) for the CERN CMS experiment.
The AMC13 is responsible for the distribution of the synchronous clock and control signals within the \mtca\ crate~\cite{Microtca:2017Online}, for the readout of the AMCs within the \mtca\ crate, and for the transmission of the data to the DAQ front-ends, which will be described in the next subsection.

\begin{figure}[htbp]
\centering
\includegraphics[width=\linewidth]{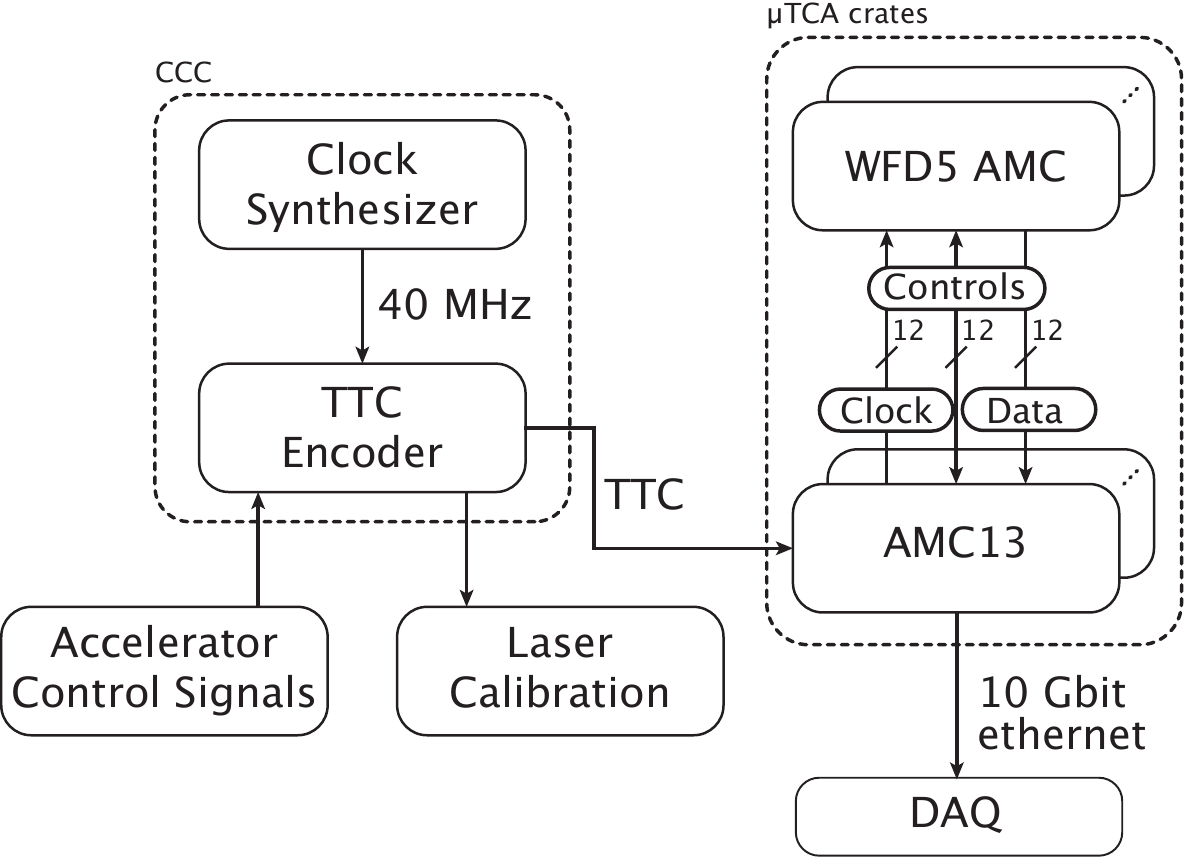}
\caption{\label{fig:beArch} The architecture of the Clock and Controls Center (CCC) for the Muon \gm\ experiment.}
\end{figure}

Figure~\ref{fig:beArch} summarizes the architecture of the clock distribution and waveform digitization systems used for the SLAC test beam run. Apart from the master reference clock, it is reflective of that employed in the Muon \gm\ experiment.
Here, an Agilent N5183A MXG Microwave Analog Signal Generator was used to provide the 40-MHz clock.

The Clock and Controls Center (CCC) takes the master 40 MHz clock and a beam-correlated signal as input and encodes both using the synchronous Timing, Trigger, and Control (TTC) protocol~\cite{Taylor:1998sx} developed for the CERN Large Hadron Collider experiments.
The TTC system uses the master clock to realize a 160 MBaud biphase mark encoder that time-division multiplexes two channels using a balanced DC-free code~\cite{TTCencode:2017Online}. 
The ``A'' channel provides a trigger to signal the start of data acquisition for a muon fill or for a laser calibration run. 
The ``B'' channel provides a variety of synchronous commands, such as clock counter resets or the type of trigger that the ``A'' channel will represent.

The TTC Encoder within the test beam's CCC system (Fig.~\ref{fig:beArch}) consists of an FC7 \mtca\ module~\cite{Pesaresi:2015qfa} outfitted with an EDA-02708 FPGA mezzanine card (FMC) to accept inputs from the clock and the accelerator controls system.
The FC7 optically transmits the encoded TTC signal to an AMC13 module in the same \mtca\ crate via a second FMC, a CERN EDA-02707.
The AMC13 distributes the TTC signal over the \mtca\ backplane to fanout FC7 modules. These modules further distribute the TTC signal through a pair of EDA-02707 FMCs to all \mtca\ client crates housing detector electronics.  The test beam had two  client \mtca\ crates, which housed the WFD5's that instrumented the PbF$_{2}$ calorimeter and the laser calibration system.

The WFD5 modules for \pb\ detectors were custom-designed for the Muon \gm\ experiment.
Each WFD5 digitization channel utilizes a Texas Instruments ADS5401 12-bit, 800-mega-samples per second (MSPS) ADC chip for the digitization.
Each channel also has a dedicated Xilinx Kintex-7 FPGA that acquires the data from the ADS5401 and buffers them in a 1-Megaword DDR3 16-bit SDRAM.
Each channel FPGA also manages the readback of its DDR3 buffer.

A sixth Xilinx Kintex-7 FPGA (the ``master'' FPGA) controls the operation of the WFD5 module and implements the external DAQ and communication links.
The master FPGA communicates with each channel FPGA over dedicated 5-Gbit serial links to provide channel control and to acquire channel data. 
The master FPGA also transfers data at 5\,Gbit/s over the \mtca\ backplane to the AMC13 with a BU-designed protocol based on an 8b/10b encoding.
An event builder on the AMC13 gathers the data from all 11 WFD5 modules from a calorimeter and transmits these data to a dedicated DAQ front-end via a 10-Gbit Ethernet link.

The 800-MHz clock for WFD5 module derives from the 40-MHz master clock recovered from the TTC signal by the AMC13 and transmitted over the backplane.
The WFD5 module provides significant flexibility in the digitization rate of each channel, with field-configurable rates ranging from 40 to 800\,MHz. 
A Texas Instruments LMK04906 clock synthesizer provides the frequency upconversion for the five channels, with a sixth output that is routed to the WFD5 front panel. For the test beam, all channels operated at the nominal 800\,MSPS.

For the SLAC electron bunches, we specified a single sample window of 560,000 ADC samples (700 $\mu$s) in length, similar to that needed for the Muon \gm\ experiment. 


\subsection{CPU-GPU hybrid system and MIDAS data acquisition}

The data acquisition system (DAQ) used was a small-scale copy of the main DAQ designed for the Muon \gm\ experiment. 
The system permitted the acquisition of dead-time-free, 700\,$\mu$s duration, continuously
digitized waveforms from the calorimeter crystals and from the laser monitors.
It was triggered by both beam and special laser-calibration events at a total trigger rate of roughly 10\,Hz.  
An event in this language corresponds to all of the hits (typically from thousands of positrons) that are accumulated during a 700\,$\mu$s muon fill in the Muon \gm\ experiment, or a sequence of laser pulses which are fired between fills. 
The concept was tested here, even though there was only one electron per event on average.

\begin{figure*}[htbp]
\centering
\includegraphics[width=0.8\linewidth]{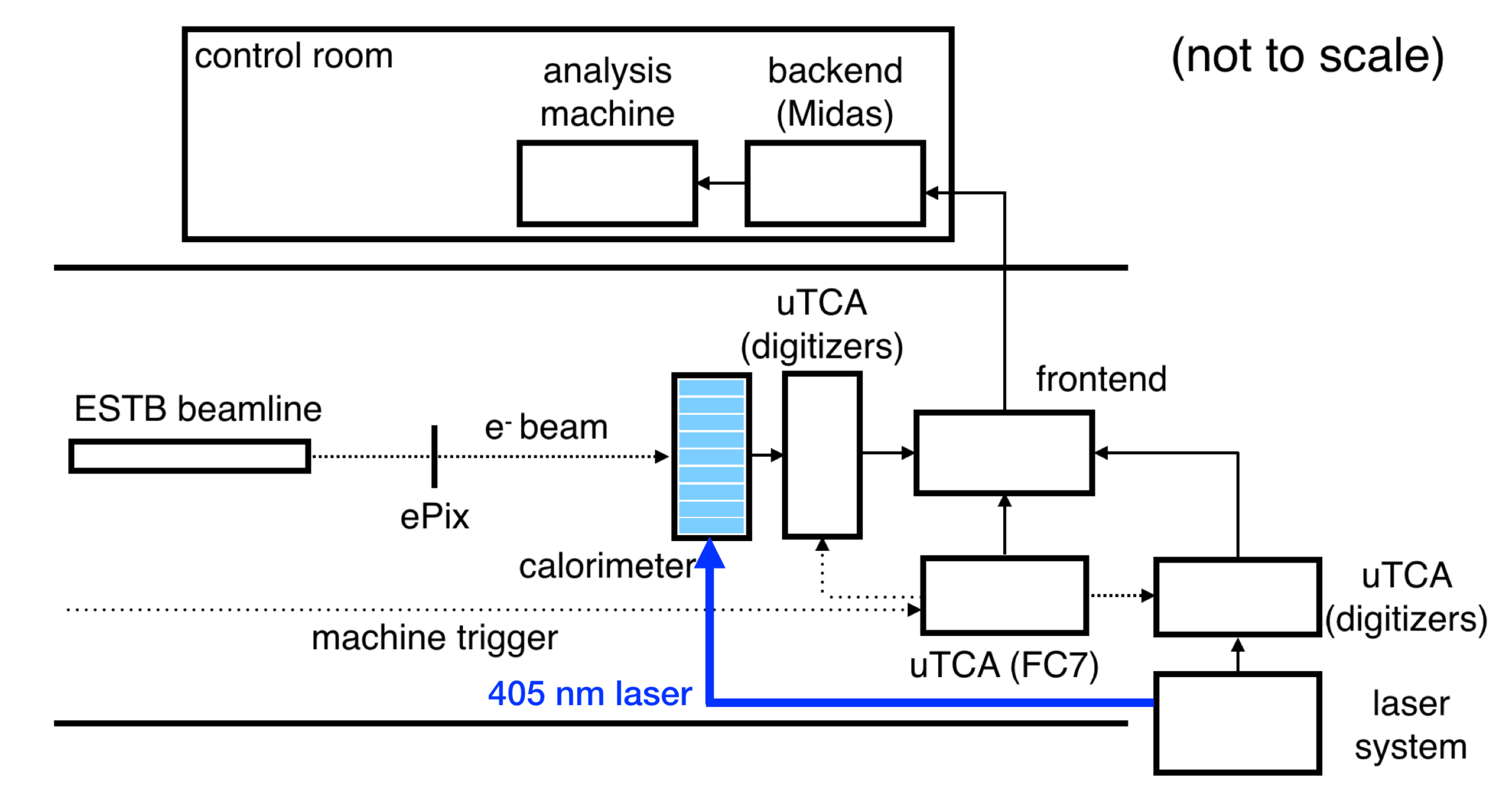}
\caption{\label{fig:overview} Overview of the experimental setup at SLAC. Electron beam from the ESTB beamline is moving from the left to right before hitting the front face of the calorimeter. The SiPM pulses resulting from electromagnetic showers, or from laser pulses, are digitized and then processed by the front-end and back-end machines to be ready for data analysis.}
\end{figure*}

The DAQ comprised a front-end computer, responsible for the readout and pre-processing of the continuously digitized waveforms from the detector systems, and a back-end computer, responsible for the event assembly, data storage, and run control.
The front-end computer reads out the raw data from the three \mtca\ crates over three point-to-point 10\,GbE fiber-optic links and pre-processed them into derived datasets using a hybrid system comprising the computer's eight-core processor and two general purpose graphical processing units (GPUs).
An overview of the data flow at the test beam is shown in Fig.~\ref{fig:overview}.
The raw data rate from the test beam electronics was approximately 2\,GByte/s and the processed data rate to the mass storage was approximately 200\,MByte/s.
In the Muon \gm\ experiment, the GPU-based pre-processing is necessary to reduce the enormous rate of continuously-digitized waveforms of approximately 20\,GByte/s to a manageable rate of approximately 200\,MByte/s of stored data.

The acquisition software is based on the \hbox{MIDAS}~\cite{Midas:2017:Online} data acquisition framework developed at PSI and TRIUMF. 
The front-end readout consists of: a {\it TCP thread} that receives and re-assembles the raw data from the AMC13 controller; 
a {\it GPU thread} that manages the GPU-based data processing into various derived datasets; 
and a {\it MIDAS thread} that handles the transfer of MIDAS-formatted events to the back-end computer event builder. 
Mutual exclusion (Mutex) locks are used to synchronize the execution of threads and ensure the integrity of data.

The GPUs were used to capture islands of ADC samples for all of the crystals in the array, regardless of their own sample values, when any of the 54 crystals exceeds a programmable threshold.
The island length includes a programmable number of pre- and post-samples around the trigger sample and is automatically extended if a second software trigger occurs during the island sample-length time period. 
The GPU processing was implemented on two NVIDIA Tesla K40 GPUs using custom CUDA kernels to parallelize the data processing over the 5776 available CUDA cores.

The MIDAS tools for event building, data storage, and run control were all hosted on the back-end computer.  
MIDAS also provided an online database (ODB) used both for saving the experimental conditions for each run and configuring the detectors, electronics, and other subsystems.

\section{Experimental Setup}
\label{sec:setup}

The experimental setup during the test beam is shown in Fig.~\ref{fig:overview}.
The calorimeter was located about 50\,cm downstream of the beamline exit window.
The electron beam size on the front face of the calorimeter, after exiting the vacuum pipe and traveling through air, was $\approx$ $0.5 \times 1.0$\,cm$^{2}$.
The mean transverse position was stable throughout the measurement period.
The calorimeter, waveform digitizers, bias control, low-voltage, and front-end computer were located inside the beam tunnel. 
The laser and monitoring system was located just outside of the tunnel, and the DAQ backend and analysis machines were located in the control room several floors above the tunnel.
This setup is representative of the distributed setup now employed at Fermilab Muon Campus.

The End Station A secondary beamline provides a well-collimated beam of electrons at a user-defined intensity, with a typical beam-pulse frequency of 5\,Hz.
In each precisely timed pulse, a Poisson-distributed number of electrons arrive.
During our normal mode of operation, the single-electron beam event was optimized at a probability of 37\%.
Energies from 2.5 to 5\,GeV were used in evaluating the performance of the calorimeter system, with each energy requiring a setup period by the operators. 
Once the beam energy was established, the rate was stable and the energy constant to better than 1\%. 
Choosing new energies required some hours of setup and the actual energy was typically accurate to 10\% of the target value. 
This limited the study of linearity by sweeping beam energies.
With these considerations, it was prudent to fix 3\,GeV as the nominal energy for the bulk of the data taking. 
It corresponds to the endpoint energy of the decay positron in the Muon \gm\ experiment.

\begin{figure}[htbp]
\centering
\includegraphics[width=\linewidth]{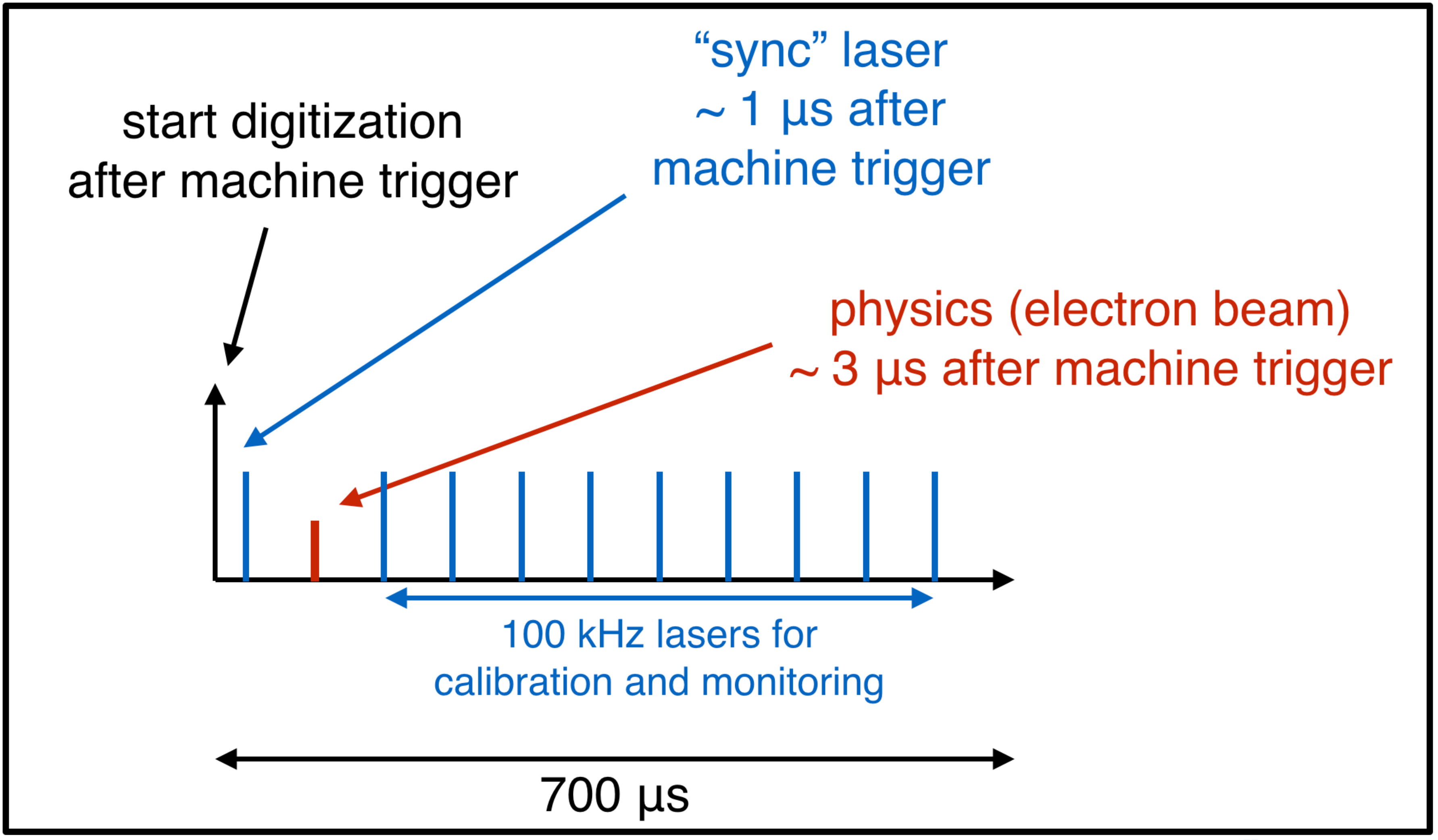}
\caption{\label{fig:eventtopology} Electron event topology at SLAC. The laser synchronization (sync) pulse arrives at $\sim 1\,\mu$s after the WFD5 begins digitization. The electron beam then hits the front face of the calorimeter 2\,$\mu$s later. Laser shots for calibration and monitoring purposes are then fired at 100\,kHz in this particular example.}
\end{figure}

Approximately 3\,$\mu$s prior to the arrival of the beam, a machine trigger was received from the SLAC accelerator control system. 
This trigger was then fed into the FC7 to be encoded together with the 40\,MHz clock. 
The FC7 then sent TTC signals to the AMC13 in its \mtca\ crate and also to the \mtca\ crates of calorimeter and laser systems, which initiated digitization of the WFD5s in both crates. 
The FC7 also sent a NIM signal to a Laser Control Board~\cite{Anastasi:2017sos} which triggered the PicoQuant laser head in a pre-configured laser firing pattern following a 1\,$\mu$s delay. 
An example pattern implemented during the test beam is shown in Fig.~\ref{fig:eventtopology}.

\section{Event reconstruction and single crystal response}
\label{sec:recon}
The offline data analysis was performed using the \textit{art}-based event processing framework~\cite{Green:2012gv,KimSiangKhawonbehalfoftheMuong-2:2017xmr} developed at Fermilab.
Data products consist mainly of fits to digitizer islands to extract pulse-integrals, fit times, and pedestals.
The fit results require energy calibration to MeV units and time-dependent gain corrections.

\begin{figure}[htbp]
\centering
\includegraphics[width=\linewidth,trim=25 0 40 0,clip]{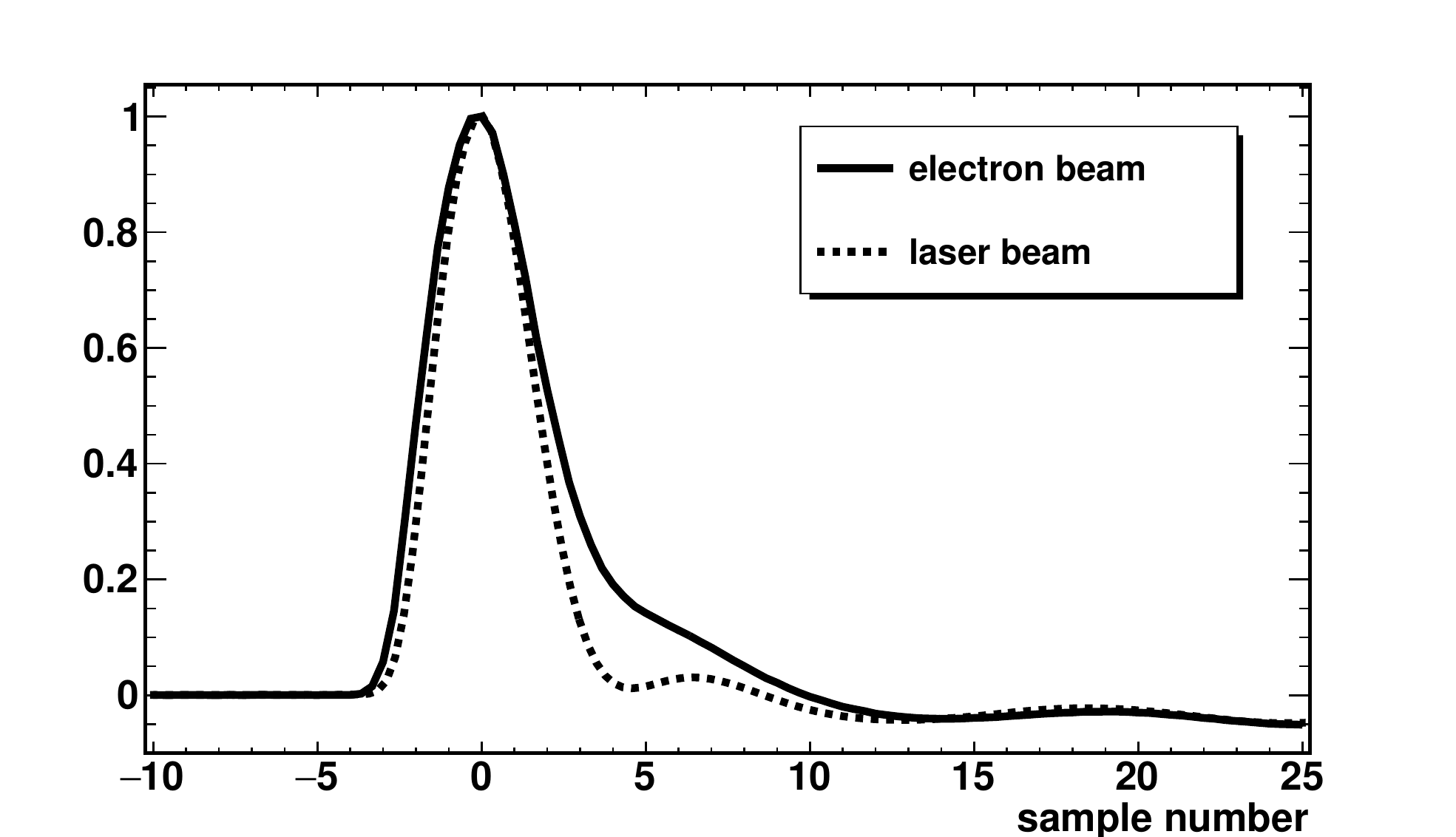}
\includegraphics[width=\linewidth,trim=25 0 40 0,clip]{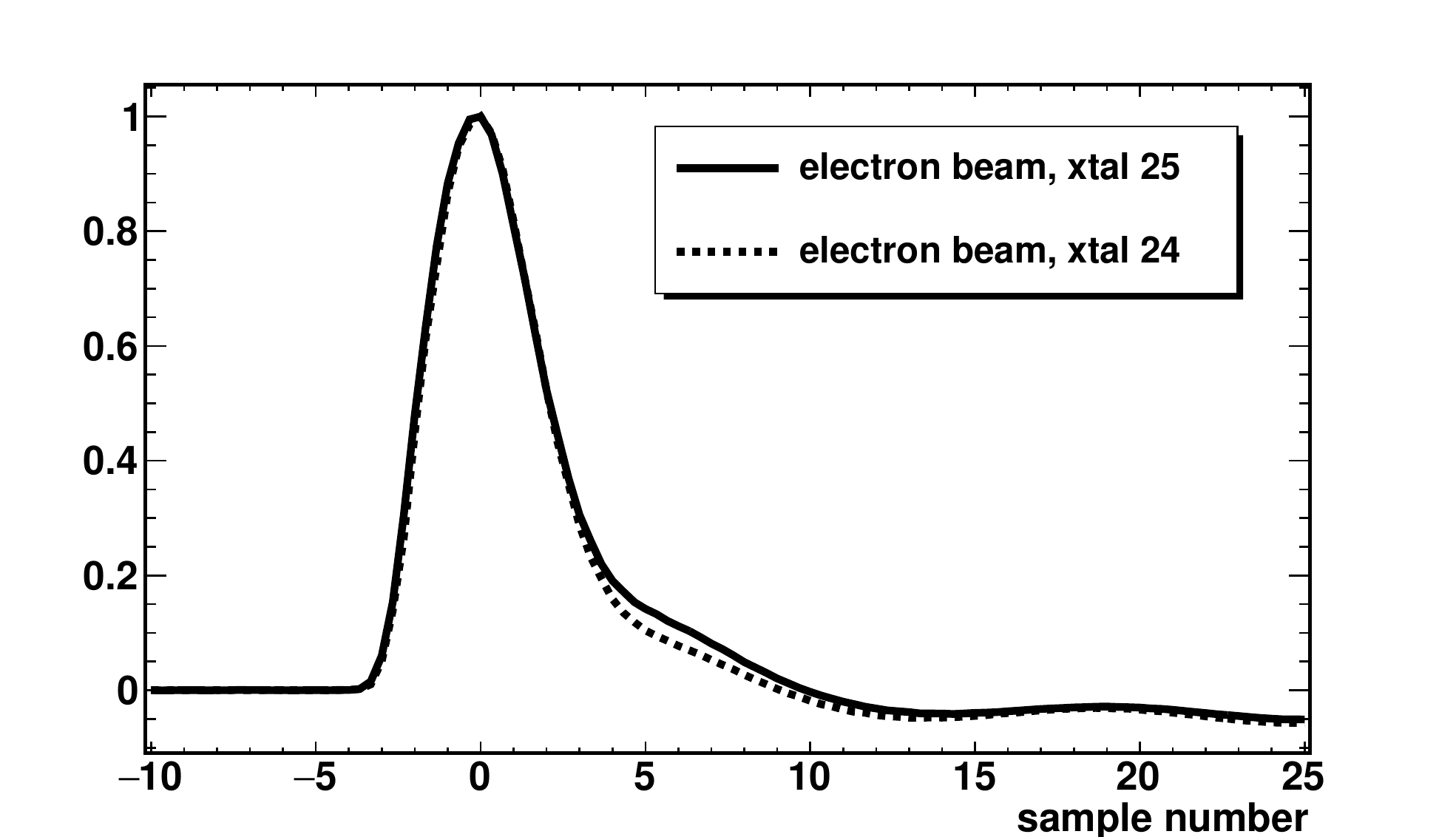}
\caption{\label{fig:beamlasertemplates} Upper: Overlay of the electron-beam template and the laser-beam template for the same crystal. Lower: Overlay of the electron-beam templates from two different crystals.}
\end{figure}

The pulse-integral is an effective measure of the number of SiPM pixels fired.
It is extracted along with the definition of the hit time from a fit to the digitized trace.
The pulse-fitter is based on custom pulse templates $T(t)$ for each individual SiPM, under the assumption that the method is robust against small fluctuations in pulse shape.
The procedure for template creation was described in Ref.~\cite{Fienberg:2014kka}.
Separate templates were built for each of the 54 crystals and for both electron showers and laser events. 
A comparison of electron-beam and laser-beam templates for the same crystal is shown in the upper panel of Fig.~\ref{fig:beamlasertemplates} while a comparison of electron-beam templates for two different crystals is shown in the lower panel.

The function used to fit the traces is:
\begin{equation}
  f(t) = s\cdot T(t-t_{0}) + P \quad .
\end{equation}
The three free parameters include an overall scale factor ($s$), the peak time ($t_{0}$) and
the pedestal ($P$). The pulse-integral is extracted as $s$.
The eigen~\cite{Eigenweb:2017Online} linear algebra library is utilized in the fitting process owing to its computing performance. 
This procedure results in a pulse-processing rate of approximately 65,000 pulses per second per CPU\footnote{Tested on a Macbook Pro 2015 with a quad-core processor of 2.5 GHz Intel Core i7}, which exceeds the expected data rate for a single calorimeter in the Muon \gm\ experiment. 
As described in \cite{Kaspar:2016ofv}, the technique permits separation of pileup events with 100\% probability for separation times larger than 5\,ns.

\subsection{SiPM pulse-shape stability under various conditions}

\begin{figure}[htbp]
\centering
\includegraphics[width=\linewidth,trim=25 0 40 0,clip]{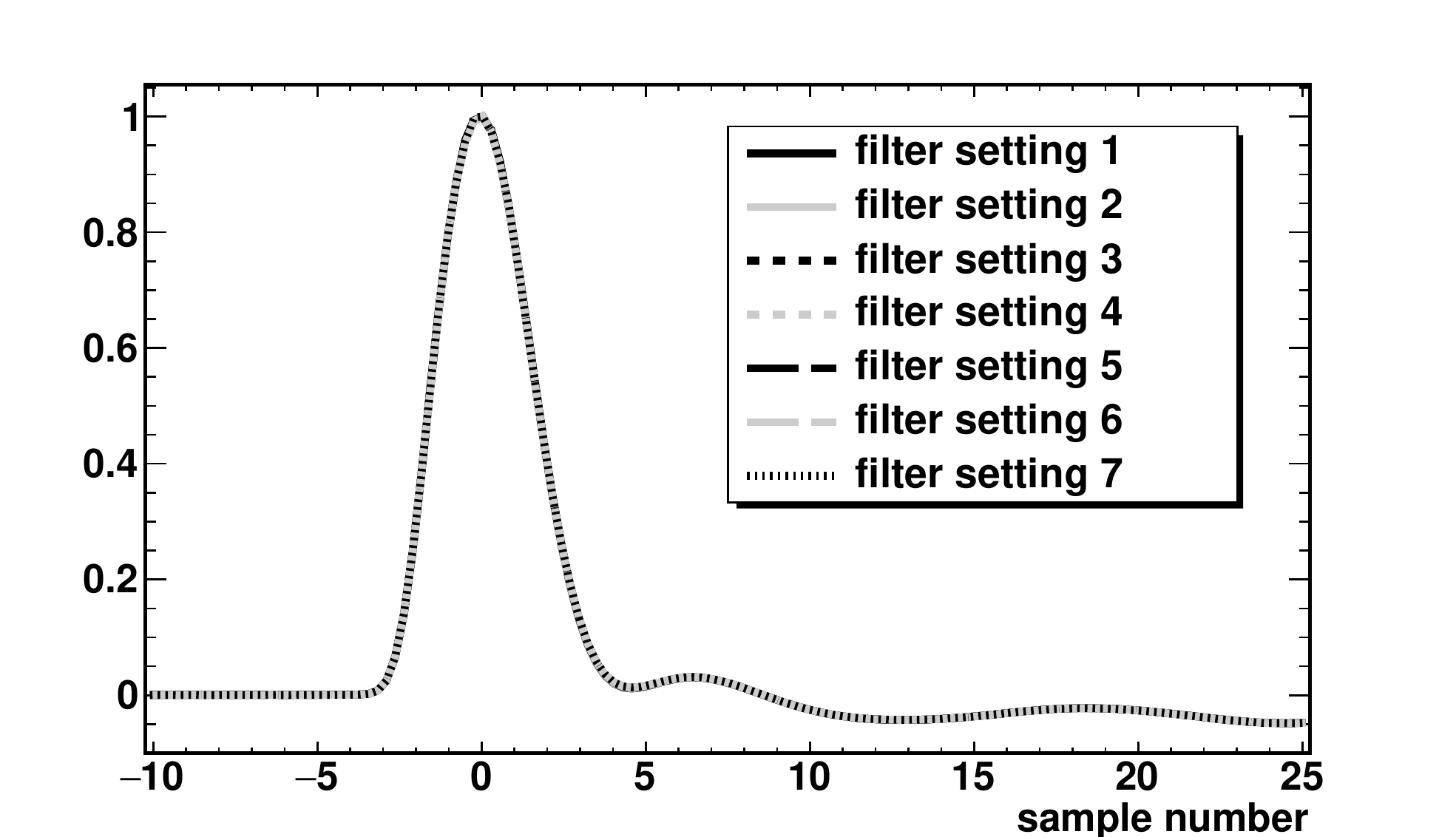}
\includegraphics[width=\linewidth,trim=25 0 40 0,clip]{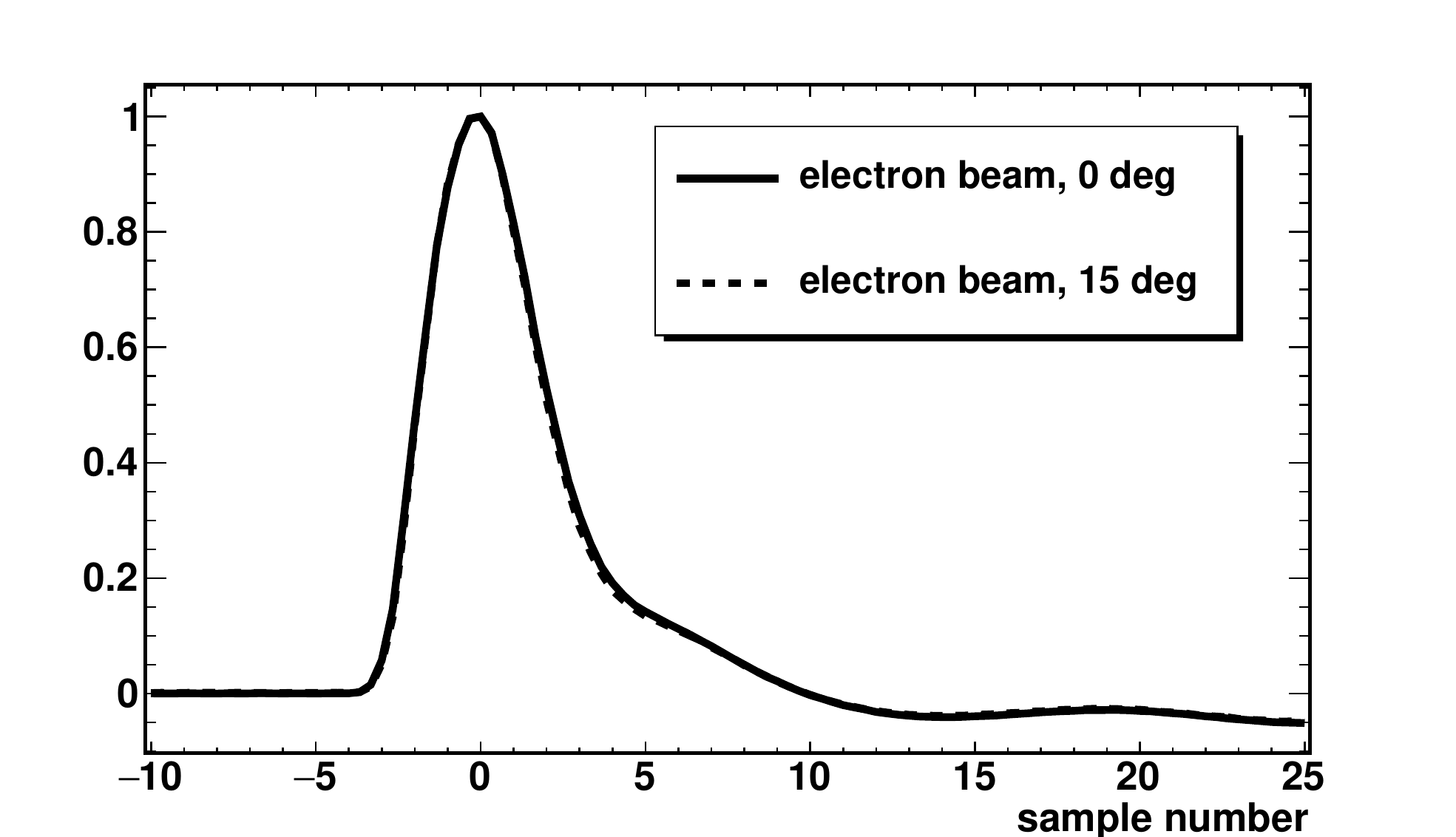}
\includegraphics[width=\linewidth,trim=25 0 40 0,clip]{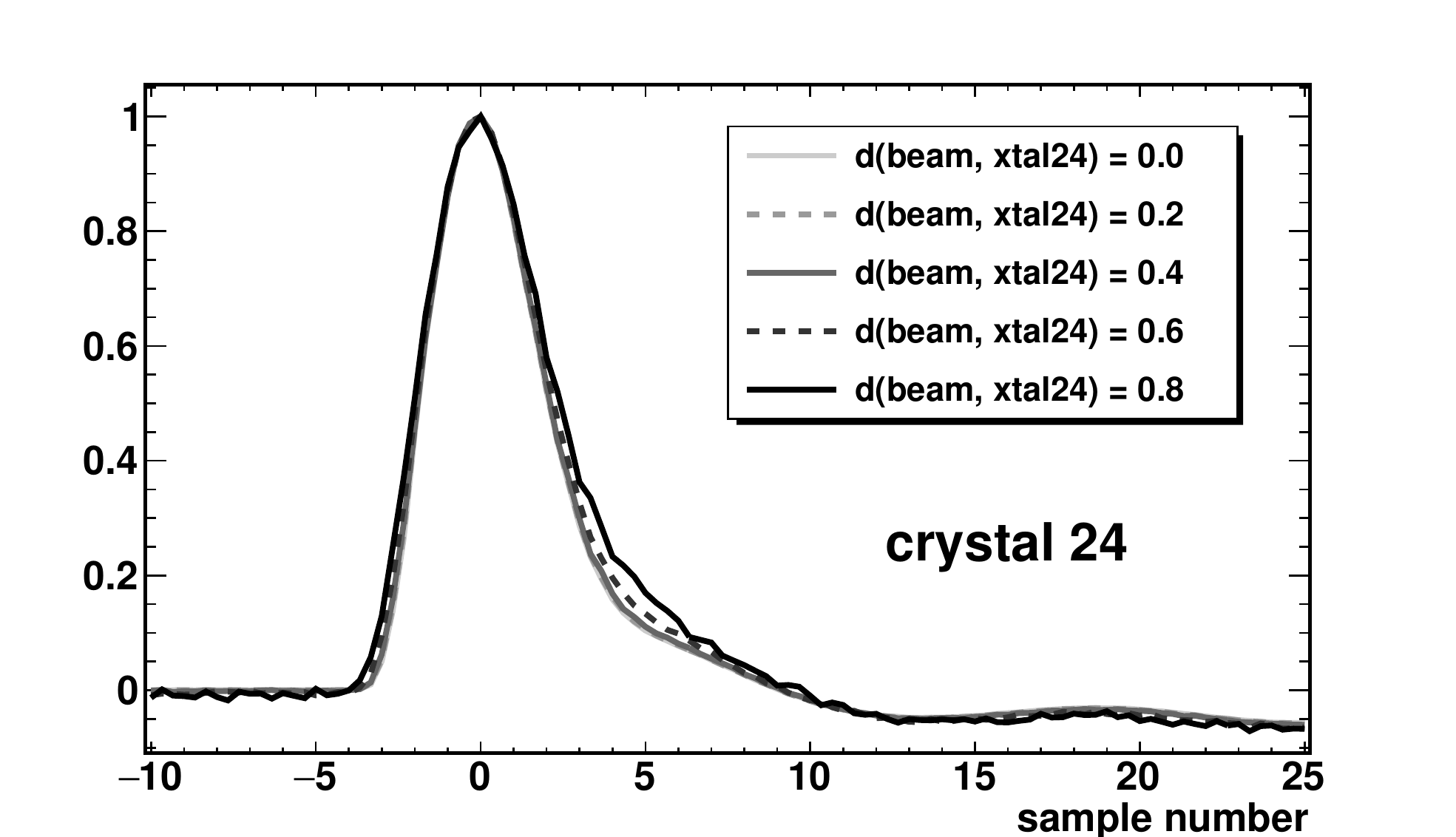}
\caption{\label{fig:pulseShape}Upper: Peak-aligned templates created from different neutral density filter wheels corresponding to different photon yields.  Middle: Fit templates created from different beam impact angles. Lower: Fit templates for various beam impact positions when the beam is moved from the center of a crystal to its direct neighbor in increments of 0.2 times a crystal width.}
\end{figure}

Template fitting relies on the stability of the SiPM pulse shape against various effects such as the number of photons (or energy), impact position, and impact angle of the beam. 
The results informed the reconstruction algorithm design.
The laser calibration system can deliver light pulses with controlled relative intensity through a rotation of its neutral density (ND) filter wheel settings. 
A series of template pulse shapes for laser events that vary in intensity from $20 - 100\%$ is shown in the upper panel of Fig~\ref{fig:pulseShape}. 
There is no noticeable difference versus the number of photons over the explored range.

The template pulse shape for different electron impact angles is a critical test for Muon \gm\ experiment, because the positron events curl into the calorimeter at energy dependent angles.
As shown in the middle panel of Fig.~\ref{fig:pulseShape}, there is a negligible difference between the 0$^{\circ}$ and 15$^{\circ}$ pulse shapes.
More noticeable is the slightly changing pulse shape versus impact position of the electron beam on the crystal face.  
As shown in the bottom panel of Fig.~\ref{fig:pulseShape}, when the electron beam is shifted from crystal 24 to its neighbor crystal 25, the region from 5 - 10 samples becomes somewhat broader. 
The impact of this effect on the energy scale and resolution is made negligible by building templates integrating over all hit positions each crystal records.

\subsection{SiPM gain equalization}
\label{subsec:gainequalization}

To establish a uniform response across the calorimeter, the SiPM gains must first be matched as closely as possible. 
Their response depends mainly on the bias voltage and the ambient temperature.
At the assembly level, the SiPMs were selected from a larger sample based on similar breakdown voltages according to the provided data sheets from Hamamatsu.

The operating temperature of the SiPMs within the calorimeter housing depends on their physical placement in the calorimeter.
A relatively continuous $\sim5^\circ$C temperature variation exists from the leftmost column to the rightmost column owing to the airflow inlet and outlet. 
The SiPMs are (approximately) arranged into groups of calorimeter columns that are then connected to one of the four independent bias supplies, which provides an average $V_{\rm{ov}}$ appropriate to the group.  
Each SiPM amplifier board contains a programmable gain amplifier (PGA) with amplification adjustable over the range 2 - 20 (80 steps), which then provides for a final adjustment.

\begin{figure}[htbp]
\centering
\includegraphics[width=\linewidth,trim=15 0 45 0,clip]{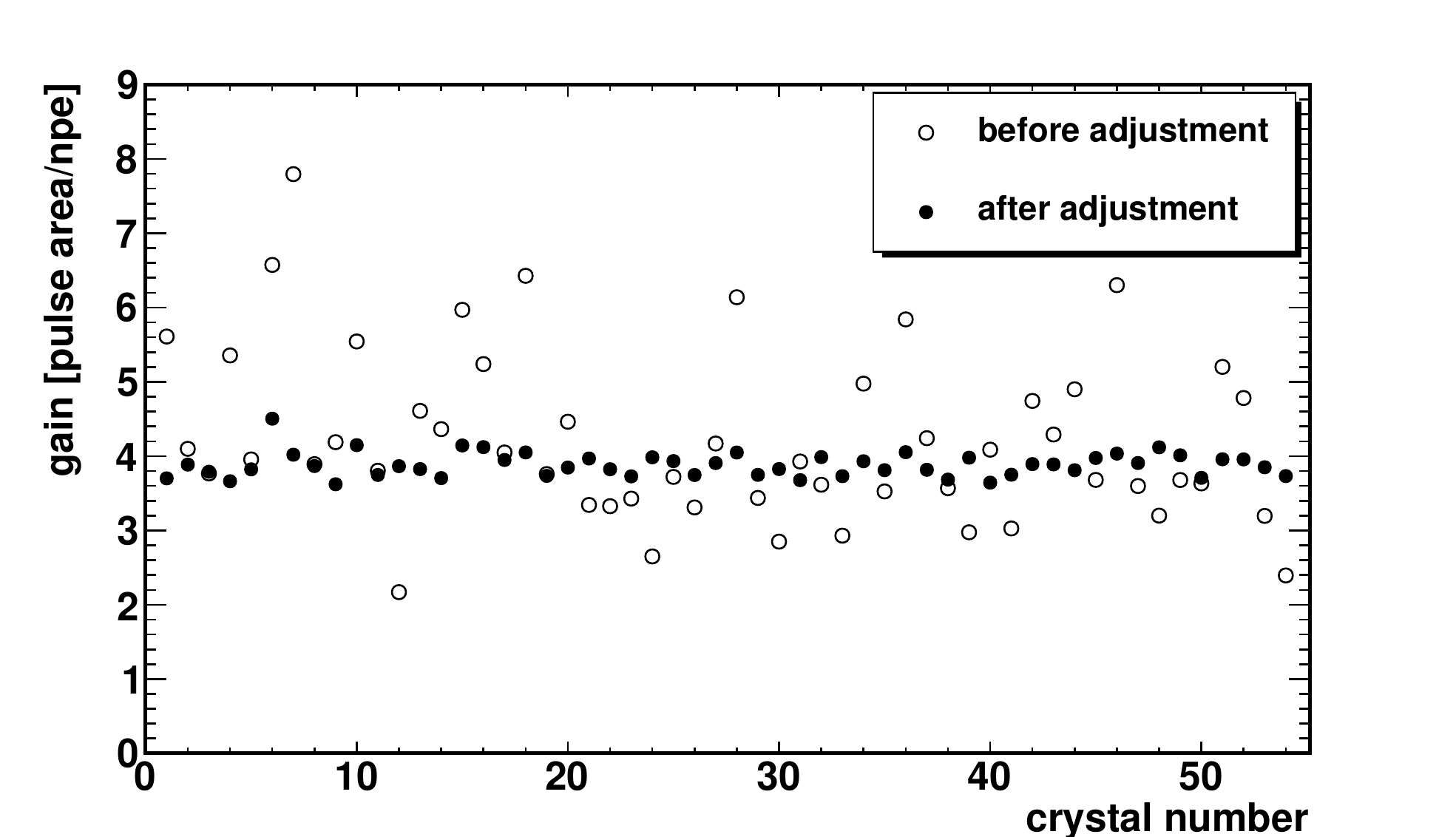}
\caption{\label{fig:calibequal} Distribution of the SiPM gain before (open circle) and after an adjustment (full circle). The uniformity of the gain after the final adjustment is 8\% in RMS.}
\end{figure}

Photon calibration is the step where the fitted pulse area can be interpreted as the number of photo-electrons~(npe), or pixels, fired for a given pulse. 
The calibration constants are obtained by employing a procedure that uses the laser and a series of runs with varying neutral density filter settings, as described in \cite{Fienberg:2014kka, Kaspar:2016ofv, Anastasi:2016luh}. 
An iterative sequence was used alternating between laser-calibration runs and adjustments to the PGA gain settings of the SiPMs and the SiPM group bias voltages. 
After several iterations, equalization of the calibration constant at the level of 8\% in RMS was achieved as shown in Fig.~\ref{fig:calibequal}. 
Further improvement is not necessary as the energy scale of each calorimeter channel does not depend solely on the SiPM gain but also on the SiPM photo-detection efficiency, the \pb\ light yield, and the \pb\ light transmission. An energy-scale equalization and calibration technique utilizing the electron beam is described in the next subsection.

\subsection{Energy scale equalization and calibration}

\begin{figure}[htbp]
\centering
\includegraphics[width=\linewidth, trim=0 0 45 0,clip]{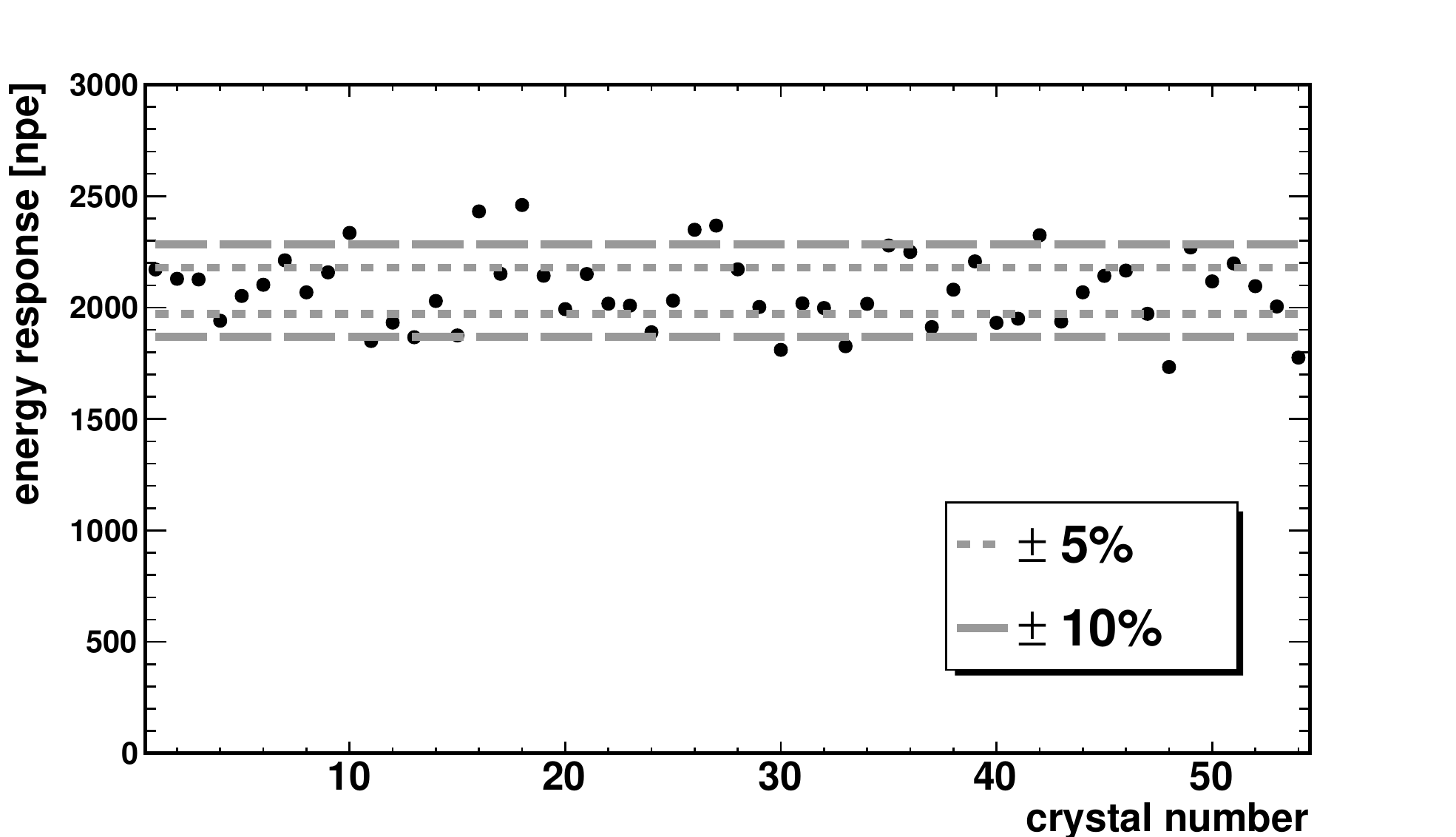}
\caption{\label{fig:energycalibration} The response in npe for each crystal in the calorimeter to a 3\,GeV electron beam impinging on the center of each crystal. A single crystal contains approximately 85\% of the shower energy for a centered electron.}
\end{figure}

In order to extract the energy calibration constants that convert npe to GeV, the 3\,GeV electron beam was aimed in turn at the center of each crystal.
A single crystal contains approximately 85\% of the shower energy, or 2.55\,GeV.
The number of photo-electrons fired for each of the crystals is shown in Fig.~\ref{fig:energycalibration}. 
The average of $\sim2100$\,npe corresponds to an energy calibration constant of $\sim0.82$\,npe/MeV, consistent with the value reported in \cite{Fienberg:2014kka}.
The accuracy of this technique is somewhat limited by the $0.5 \times 1.0$\,cm$^{2}$ beam spot and the precision with which the calorimeter could be aligned to the center of the beam.

\subsection{Maintaining SiPM gain calibration over time}

During the test beam run, laser calibration filter wheel scans were performed on average every three hours to establish the stability of the SiPM gain (conversion constant of npe to pulse integral).
To correct for the SiPM gain drift between two consecutive filter wheel scans, which is expected owing to environmental temperature fluctuations, a different procedure was employed. 
A hybrid fill structure was created as shown in Fig.~\ref{fig:eventtopology}.
The light pulses that strike a SiPM are, in time order: the laser sync pulse, the Cherenkov lights from an electromagnetic shower, and a series of approximately 70 in-fill laser pulses.
The gain drift correction is made with respect to a reference run, where the average laser pulse area for each crystal is $L_{0}$. 
The average of the SiPM response to the series of in-fill laser pulses for each fill $i$ is $\langle L_{i}\rangle$.
A gain correction factor $C_{\rm{SG}} = L_{0}/{\langle L_{i}\rangle}$ is established for each fill.

To correct for intrinsic fluctuations of the laser, information from source monitors is used similarly.
The average response of PIN1 and PIN2 during the reference run is $S_{0}$, and the average of the in-fill laser pulses measured by source PIN1 and PIN2 is $\langle S_{i}\rangle$.
The laser fluctuation correction factor $C_{\rm{LF}} = \langle S_{i}\rangle/S_{0}$ is then applied to the gain corrected crystal hit.

\begin{figure*}[htbp]
\centering
\includegraphics[width=.49\linewidth, trim=0 0 40 0,clip]{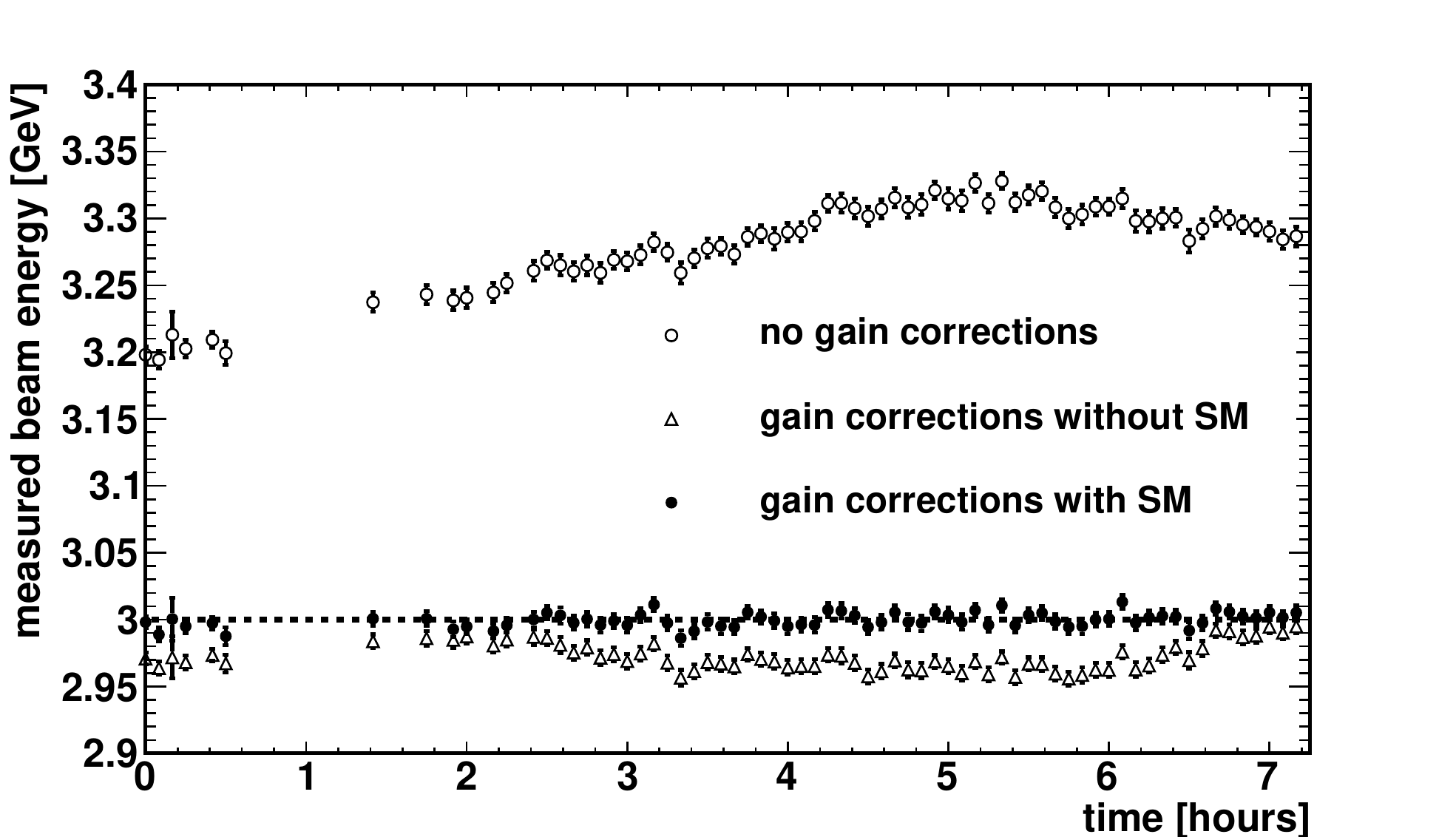}
\includegraphics[width=.49\linewidth, trim=0 0 40 0,clip]{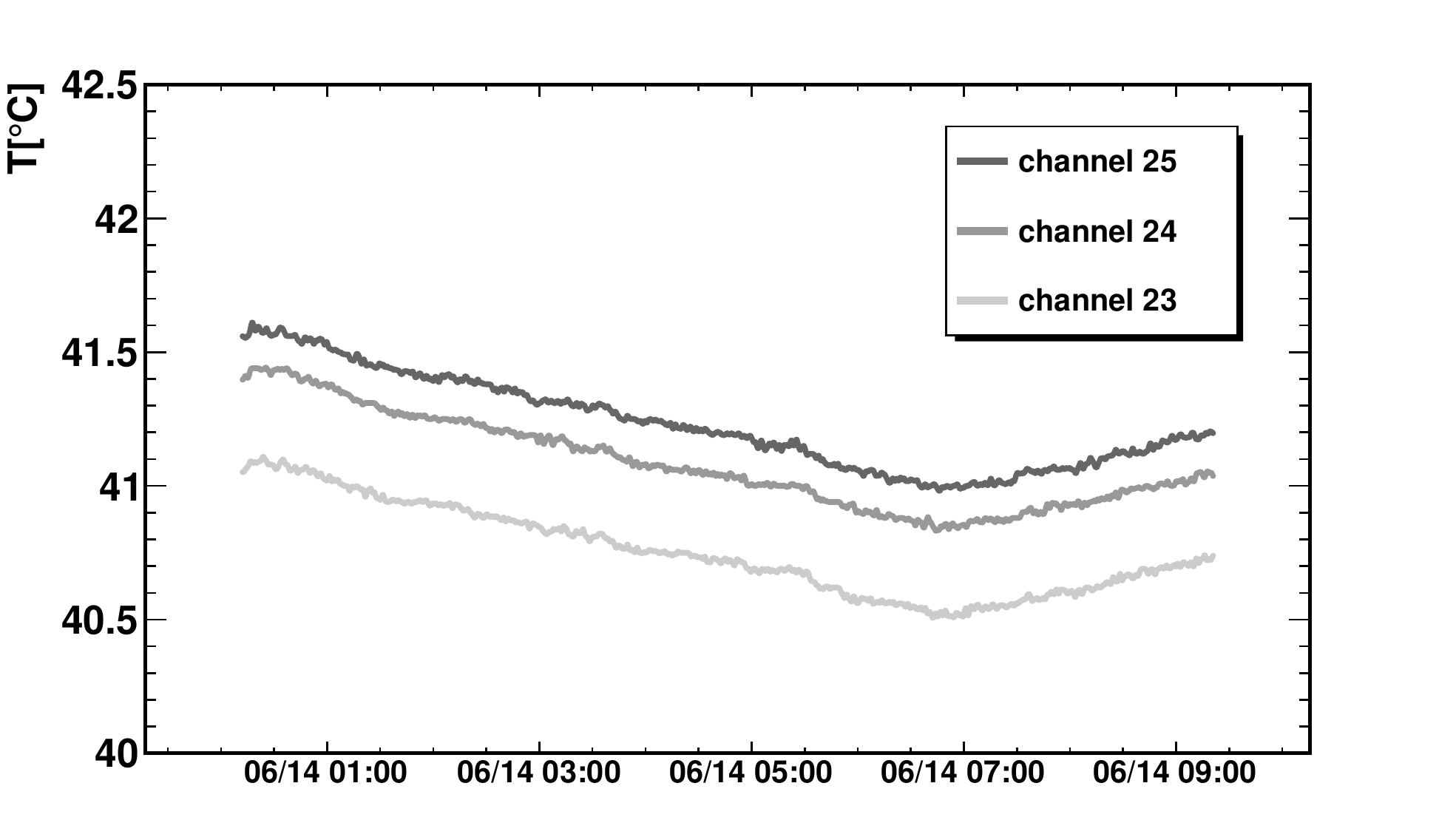}
\caption{\label{fig:energystability}Left: Stability of the energy scale of the reconstructed electron objects over roughly 7 hours. Right: Temperature distribution of three of the SiPMs over the corresponding period.  Note the relationship between temperature drift and uncorrected gain drift.}
\end{figure*}

The resultant energy stability is shown in the left panel of Fig.~\ref{fig:energystability}, while the temperature of the corresponding period is shown in the right panel. 
The drop in temperature coincides with an increase in the energy scale.
In the actual Muon \gm\ experiment, a laser pulse sequence is fired between normal muon storage ring fills.

\subsection{Time alignment of digitizer channels}
\label{subsec:timealign}

\begin{figure}[htbp]
\centering
\includegraphics[width=\linewidth,trim=5 0 45 0,clip]{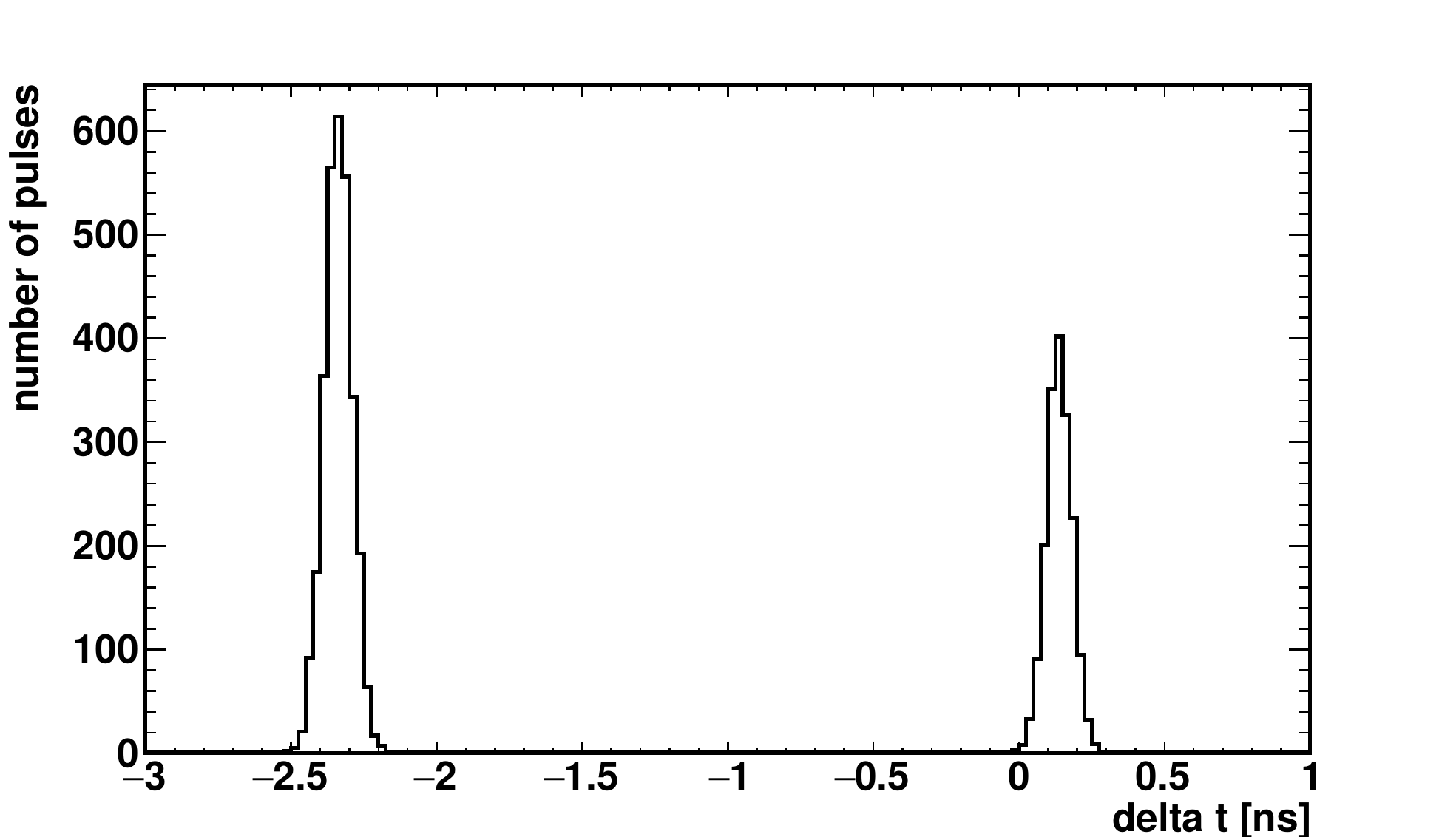}
\caption{\label{fig:syncpulse} The time difference distribution between channel 12 and 44
for the same laser event.}
\end{figure}

As multiple WFD5 modules and multiple channels in a WFD5 were used, it was necessary to align these channels in time to enable accurate hit clustering and to optimize the rejection of multi-particle pileup. 
To align all the digitizer channels within a \mtca\ crate, the laser sync pulse, fired at the beginning of each trigger, was used. 
The pulse is distributed to all 54 crystals at the beginning of each event. 
For example, the time difference distribution of the laser sync pulse between channel 12 and 44 is shown in Fig.~\ref{fig:syncpulse}.
Two narrow distributions exist, and both of them are not centered around zero. 
The former is owing to the fact that as the master TTC clock is running at 40\,MHz while the digitizers are running at 800\,MHz, there can be small differences in the exact digitizer clock tick on which a given ADC will begin digitizing.
Because the ADS5401 ADCs always send data from the ``odd internal ADC" first, we see a clear separation of 2.5\,ns or 2 clock ticks  (1 clock tick = 1.25\,ns) as a result.
The non-zero mean value is due to the length difference in light distribution fibers. The width of the distribution will be characterized in the timing resolution section.

\subsubsection{Hit clustering}

The hit clustering algorithm used at SLAC is based solely on time partitioning, whereby all hits on a given island within a time separation of $\Delta T$ are grouped. 
The parameter $\Delta T$ varied depending on the analysis type. 
The cluster time is given by the hit time of the highest-energy crystal within a cluster.

\begin{figure}[htbp]
\centering
\includegraphics[width=\linewidth]{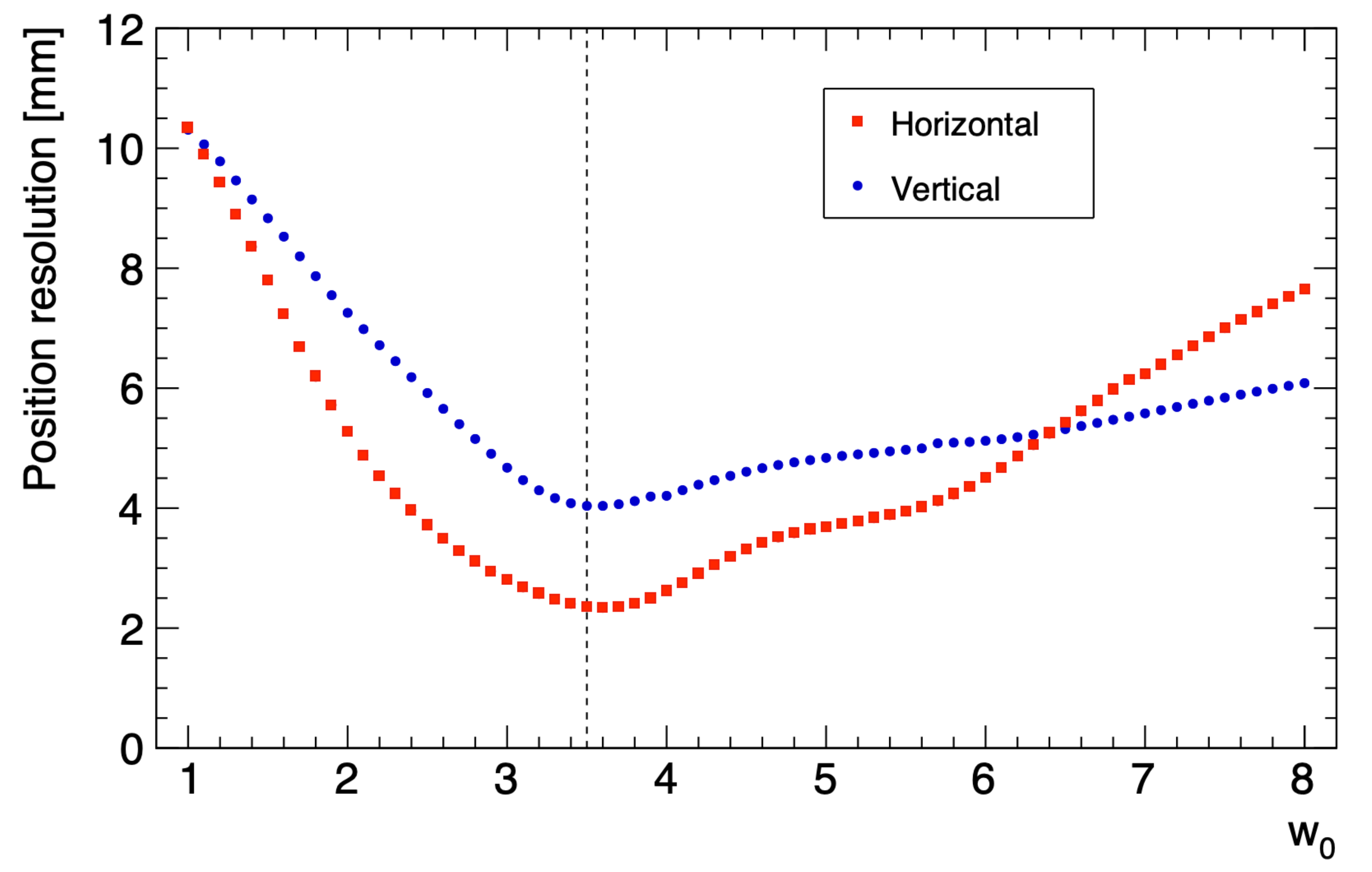}
\caption{\label{fig:final_resolution_w0} The position resolution using the logarithmic method as a function of the weighting parameter $w_0$.  The minimum at $w_0 = 3.5$ is taken at the optimization.}
\end{figure}

\begin{figure}[htbp]
\centering
\includegraphics[width=\linewidth]{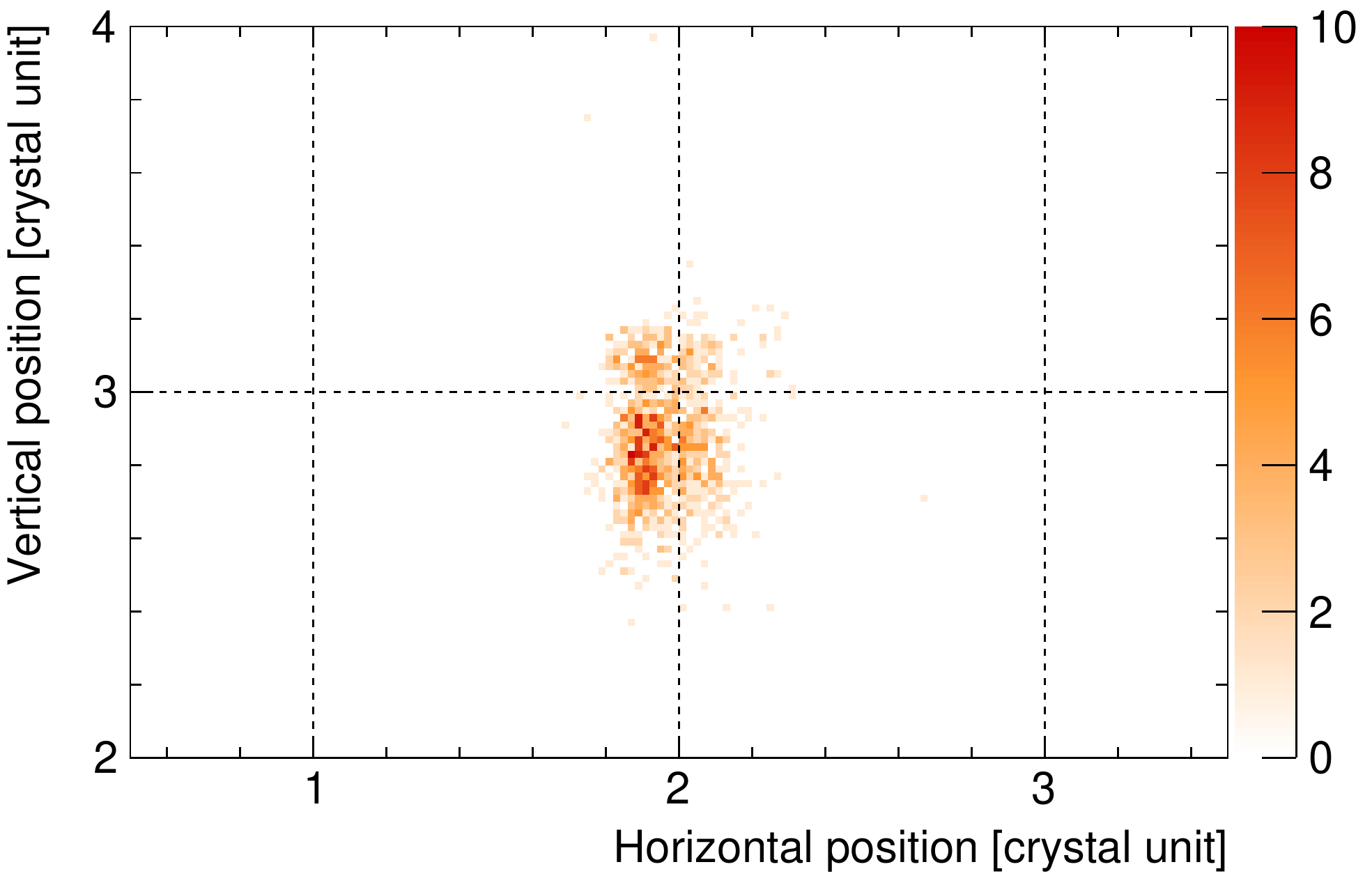}
\caption{\label{fig:paper_beam_spot} The electron beam size on the calorimeter front face as determined by the position reconstruction algorithm and verified with a finely pixelated silicon detector.}
\end{figure}

The electron's incident position is reconstructed using a center-of-gravity method with logarithmic weights~\cite{Awes:1992yp}. 
This weighting accounts for the exponential falloff of the energy deposited in each crystal across the calorimeter.  It is calculated by
\begin{equation}
(x,y) = \left(\frac{\sum_{i} w_{i} \cdot  x_{i}}{\sum_{i} w_{i}}, \frac{\sum_{i} w_{i} \cdot  y_{i}}{\sum_{i} w_{i}}\right)
\end{equation}
where $x_i$ and $y_i$ are the coordinates of the crystal column or row $i$, respectively, and $w_i$ is its logarithmic weight given by
\begin{equation}
w_{i} = \mathrm{max} \left \{ 0, \left(w_{0} + \log{\frac{E_{j}}{\sum_{j}E_{j} }} \right)    \right\}~.
\end{equation}
with $w_0$ being a free parameter that sets the relative importance of shower tails in the weighting.  
As shown in Fig.~\ref{fig:final_resolution_w0}, the optimal value of $w_0 = 3.5$ was determined by minimizing the position resolution, which agrees with Geant4 simulations.
The electron beam size, shown in Fig.~\ref{fig:paper_beam_spot} is found to be approximately twice as wide vertically than horizontally.  
It was verified with an ePix $20\times20$ mm silicon detector installed between the beamline and the calorimeter.

\section{Performance of the calorimeter system}
\label{sec:performance}

This section describes the performance of the calorimeter system as a whole in response to the electron beam. 
In all cases, the times or energies represent those of the shower clusters.
In general, the findings using our full calorimeter with black-wrapped crystals, glued-on SiPMs, optimized SiPM pulse-summing circuits, and the \gm\ custom digitizers, are equal or better than those published earlier by us using prototype devices~\cite{Fienberg:2014kka}.

\subsection{Timing resolution}

The timing measurement of the Muon \gm\ calorimeters is driven by three factors: (a) the transit time of the SiPM device, the electronics amplification circuit, and the digitizer electronics; (b) the statistical fluctuation of photons triggering SiPM pixels; and (c) the electromagnetic shower profile fluctuation, the Cherenkov light emission, and the light propagation time. 
Quantitative analyses of (a) and (b) will be given, along with typical results for (c).

\begin{figure}[htbp]
\centering
\includegraphics[width=\linewidth, trim=0 0 40 0,clip]{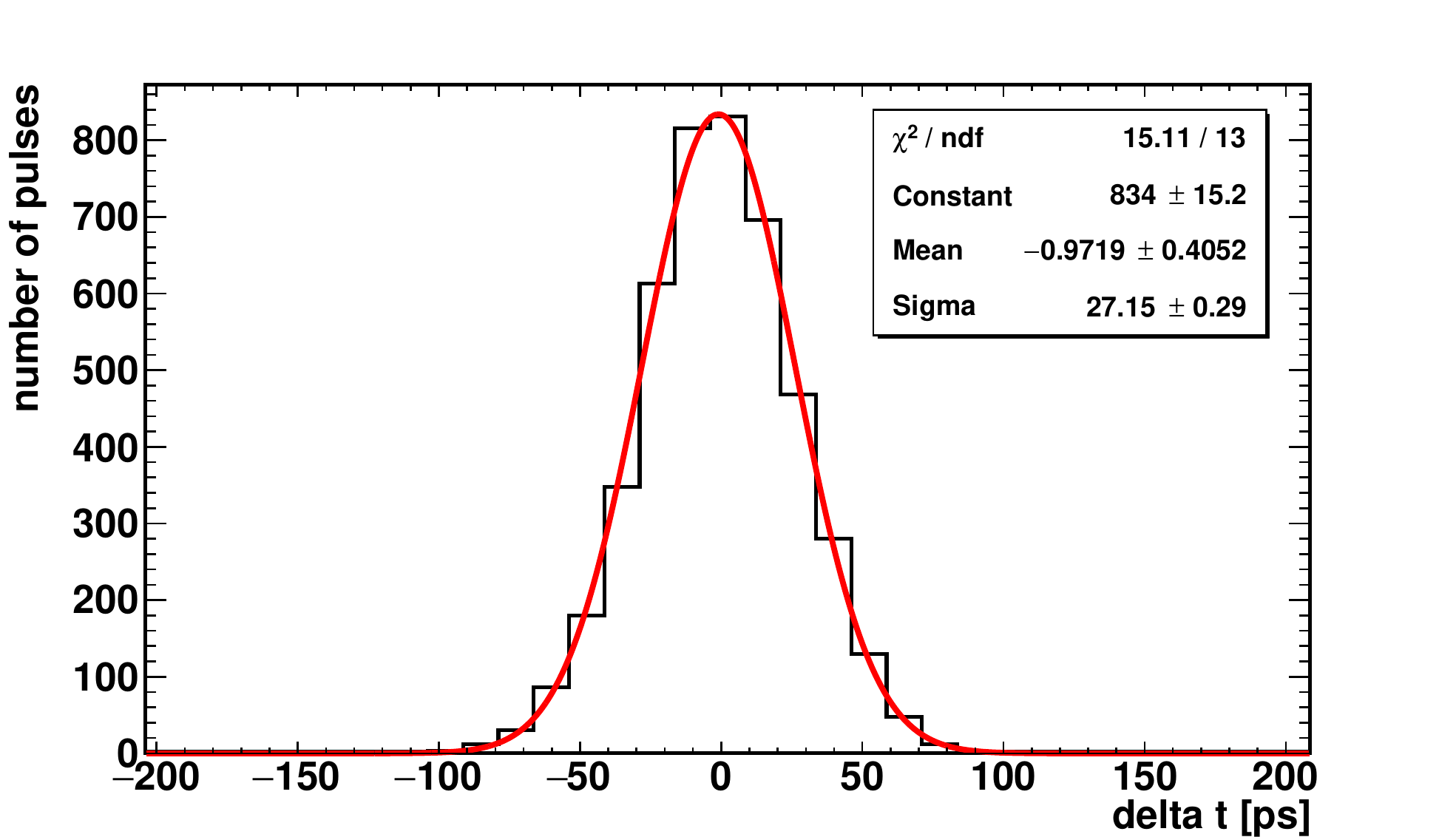}
\caption{\label{fig:timeres1} The time difference distribution between two WFD5 channels for the same laser event.}
\end{figure}

The timing resolution from (a) and (b) can be extracted using laser events that arrive at all SiPMs at about the same time.
The time difference distribution between two WFD5 channels of the same laser event is shown in Fig.~\ref{fig:timeres1}. 
The solid line is a Gaussian fit to the distribution and the standard deviation is 27\,ps for this particular set of WFD5 channels. 
The energy dependence of this timing resolution is studied for two different pairs: crystals read out by the same, or different, WFD5s.

\begin{figure}[htbp]
\centering
\includegraphics[width=\linewidth, trim=0 0 40 0,clip]{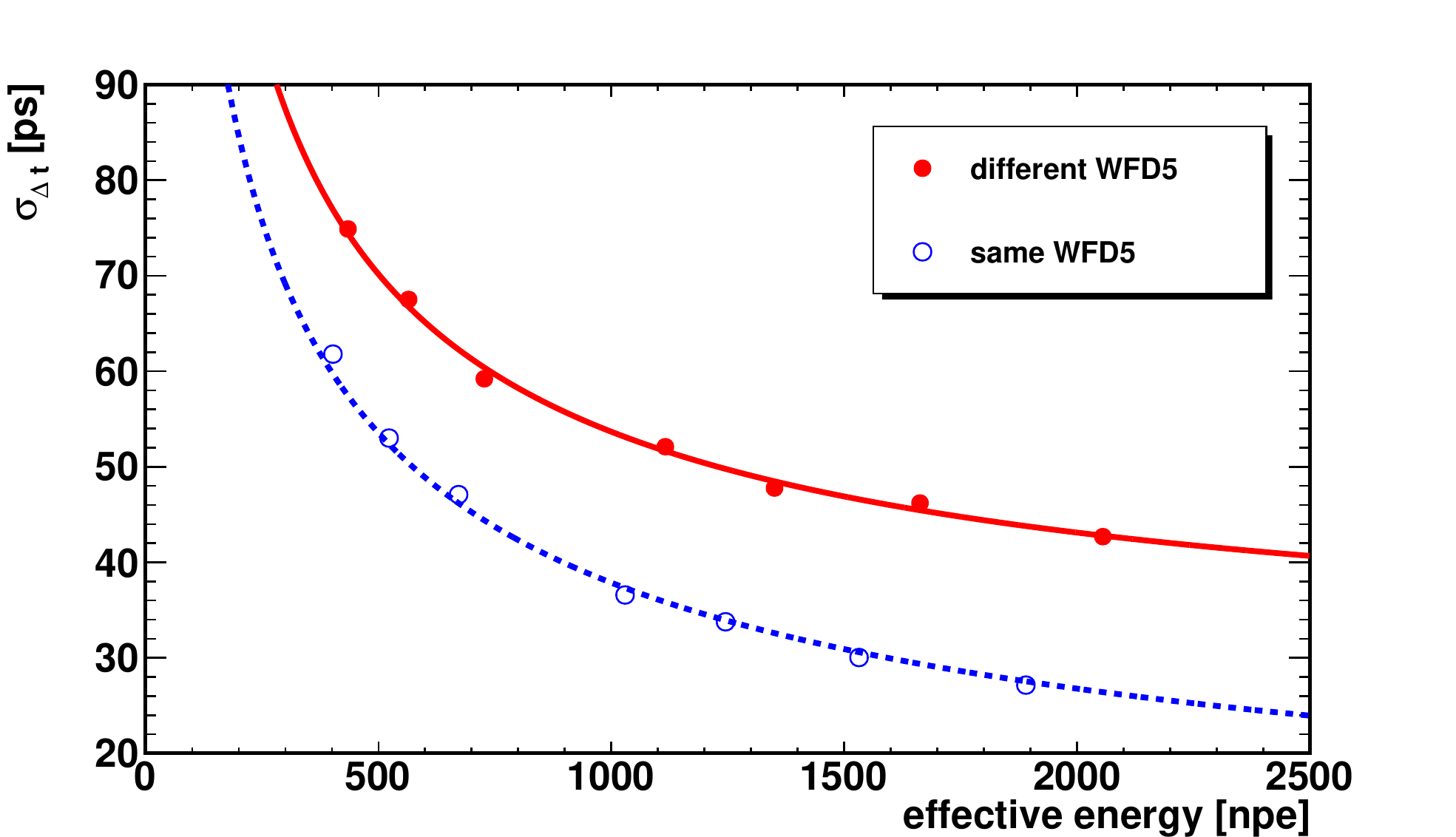}
\caption{\label{fig:timeres2} Standard deviation of the time difference distribution versus effective energy for laser events read out by the same, or different, WFD5s. A single channel resolution can be obtained by using the dotted line and scaling the vertical axis by $1/\sqrt{2}$.}
\end{figure}

The standard deviation of the time difference distribution is plotted against energy in Fig.~\ref{fig:timeres2}. 
The difference in the timing resolution between ``different WFD5" and ``same WFD5" is attributed to the timing jitter of the 800-MHz clock between two different WFD5s. 
The data points from each pair are fitted using
\begin{equation}
\sigma_{\Delta t}(E_{\rm{eff}}) = \sqrt{ 2C_{\rm{T}}^2 + \frac{S_{\rm{T}}^2}{E_{\rm{eff}}/\sigma_{\rm{n}}}  }
\end{equation}
where $C_{\rm{T}}$ is the constant term, $S_{\rm{T}}$ the stochastic term, $\sigma_{\rm{n}}$ the noise term, and $E_{\rm{eff}}$ the effective energy term defined by $E_{\rm{eff}}=E_{1}E_{2}/\sqrt{(E^{2}_{1}+E^{2}_{2})/2}$, where $E_1$ and $E_2$ are measured in the unit of npe. 
The use of $E_{\rm{eff}}$ is motivated by the fact that there is a variation in laser intensity among calorimeter channels. 
The value of $\sigma_{n}$ is fixed at 5\,npe based on digitizer noise level. 
The $C_{\rm{T}}$ term, which is mainly due to the timing jitter, is negligible for timing difference between two channels of the same WFD5. 
For two channels read out by different WFD5s, $C_{\rm{T}}$ is of the order of 25\,ps. 
On the contrary, the $S_{\rm{T}}$ term, which is related to the photo-statistics, is consistent between the two groups having an average value of approximately 600\,ps.

\begin{figure}[htbp]
\centering
\includegraphics[width=\linewidth, trim=0 0 45 0,clip]{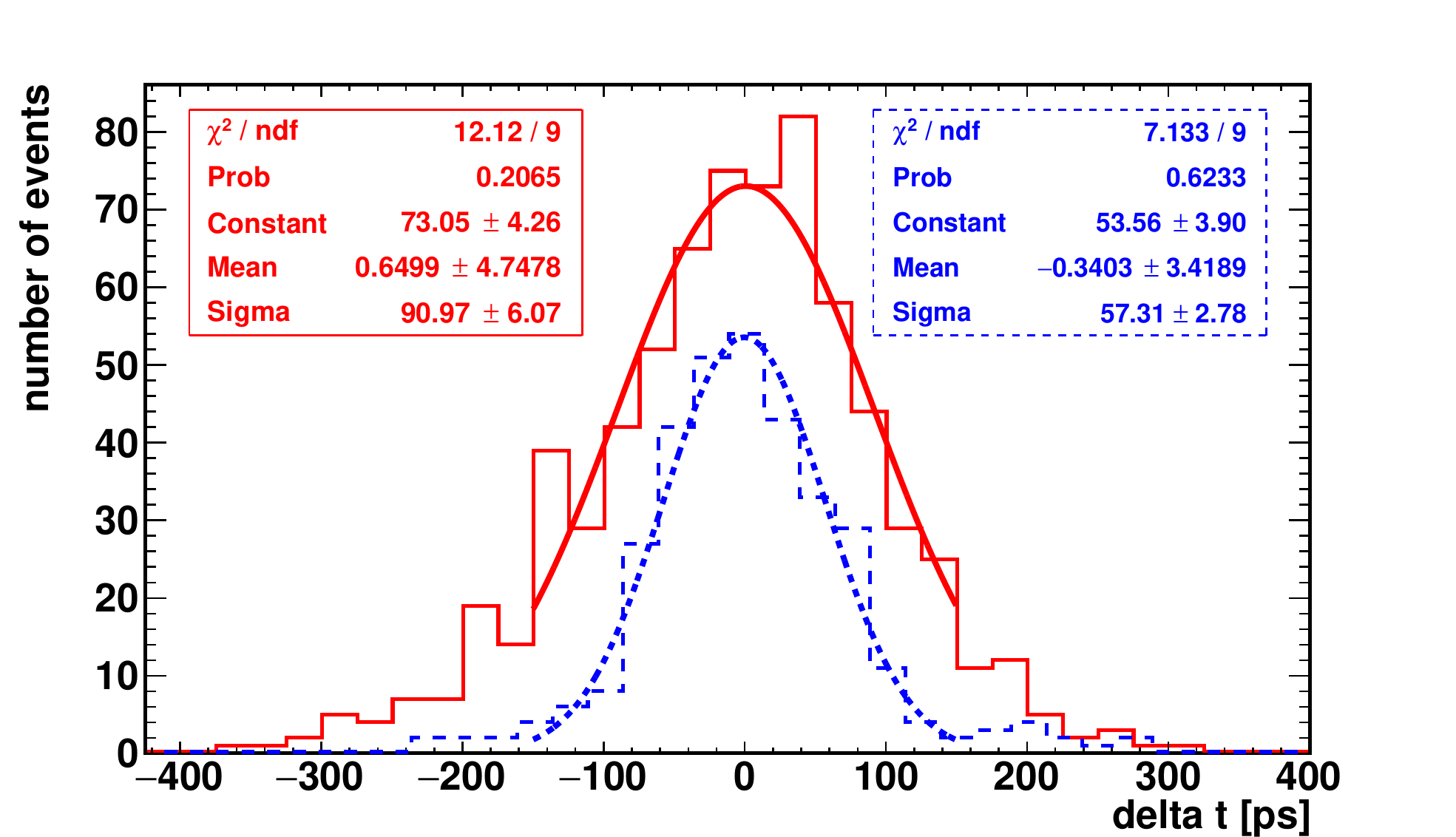}
\caption{\label{fig:dtBeam2425}Distribution of the time difference between two channels from the same WFD5 (dotted histogram) and different WFD5 (solid histogram), for the same electron event. The solid (dotted) line is a Gaussian fit to the solid  (dotted) histogram.}
\end{figure}

To extract the timing resolution of an electromagnetic shower, we studied the time difference between two calorimeter channels for electron events. 
The distribution of the time difference is shown in Fig.~\ref{fig:dtBeam2425}, for two channels from the same WFD5 and different WFD5. 
The single channel timing resolution for a 3\,GeV electron event is $57/\sqrt{2} \sim 40$\,ps (dotted line). 
A systematic and precise way of quantifying this resolution is difficult due to the broad beam size and uncertainty in the beam hit position.

\subsection{Energy resolution and linearity}

\begin{figure}[htbp]
\centering
\includegraphics[width=\linewidth, trim=0 0 40 0,clip]{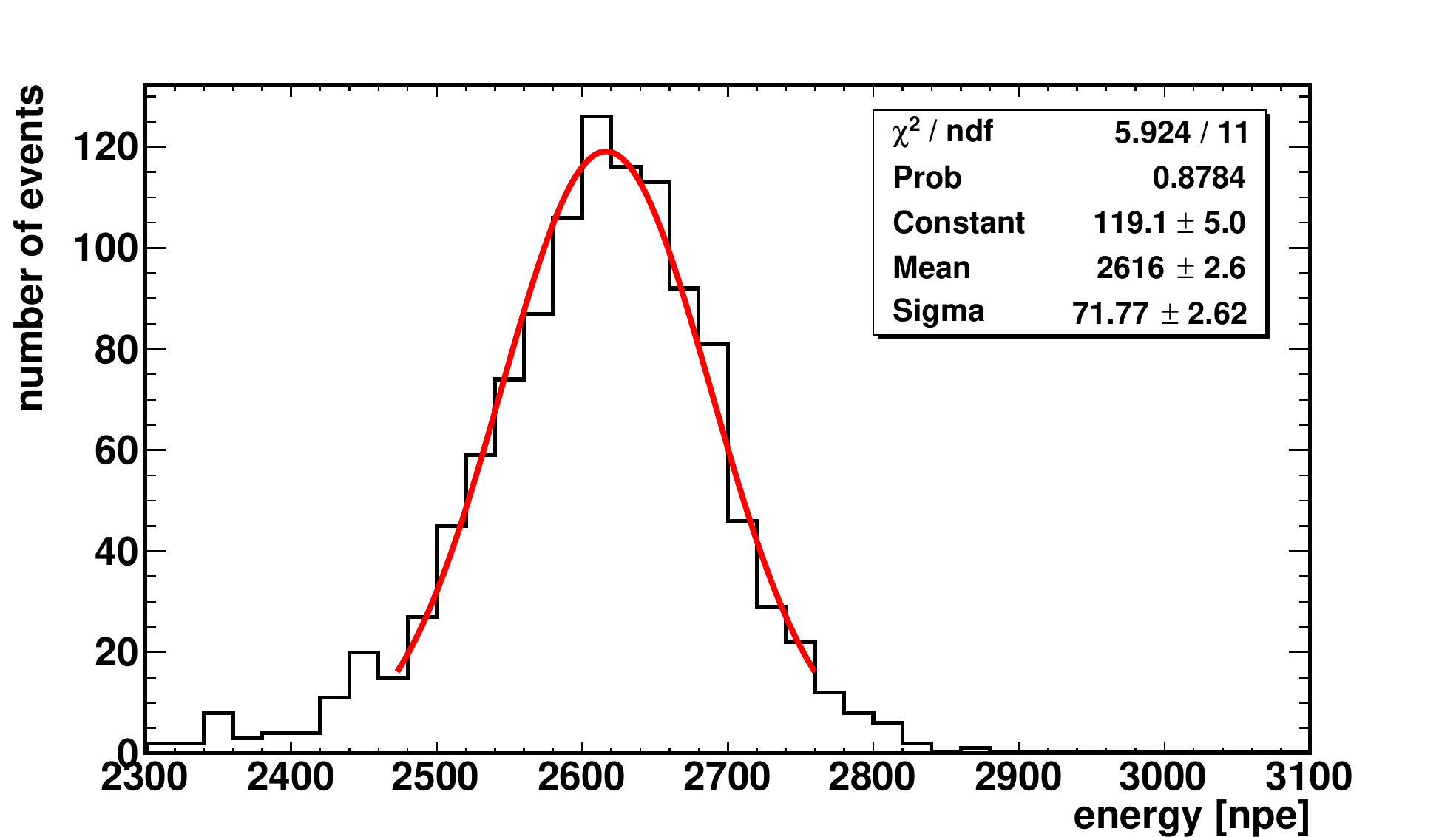}
\caption{\label{fig:caloedist}Measured cluster energy distribution for 3\,GeV electrons aimed at a typical crystal.}
\end{figure}

Figure~\ref{fig:caloedist} shows the energy distribution when a 3\,GeV beam is centered at a particular crystal.
The solid line is a Gaussian fit within $\pm 2\sigma$ around the mean value. The fitted energy resolution here is 2.7\%.

\begin{figure}[htbp]
\centering
\includegraphics[width=\linewidth, trim=0 0 40 0,clip]{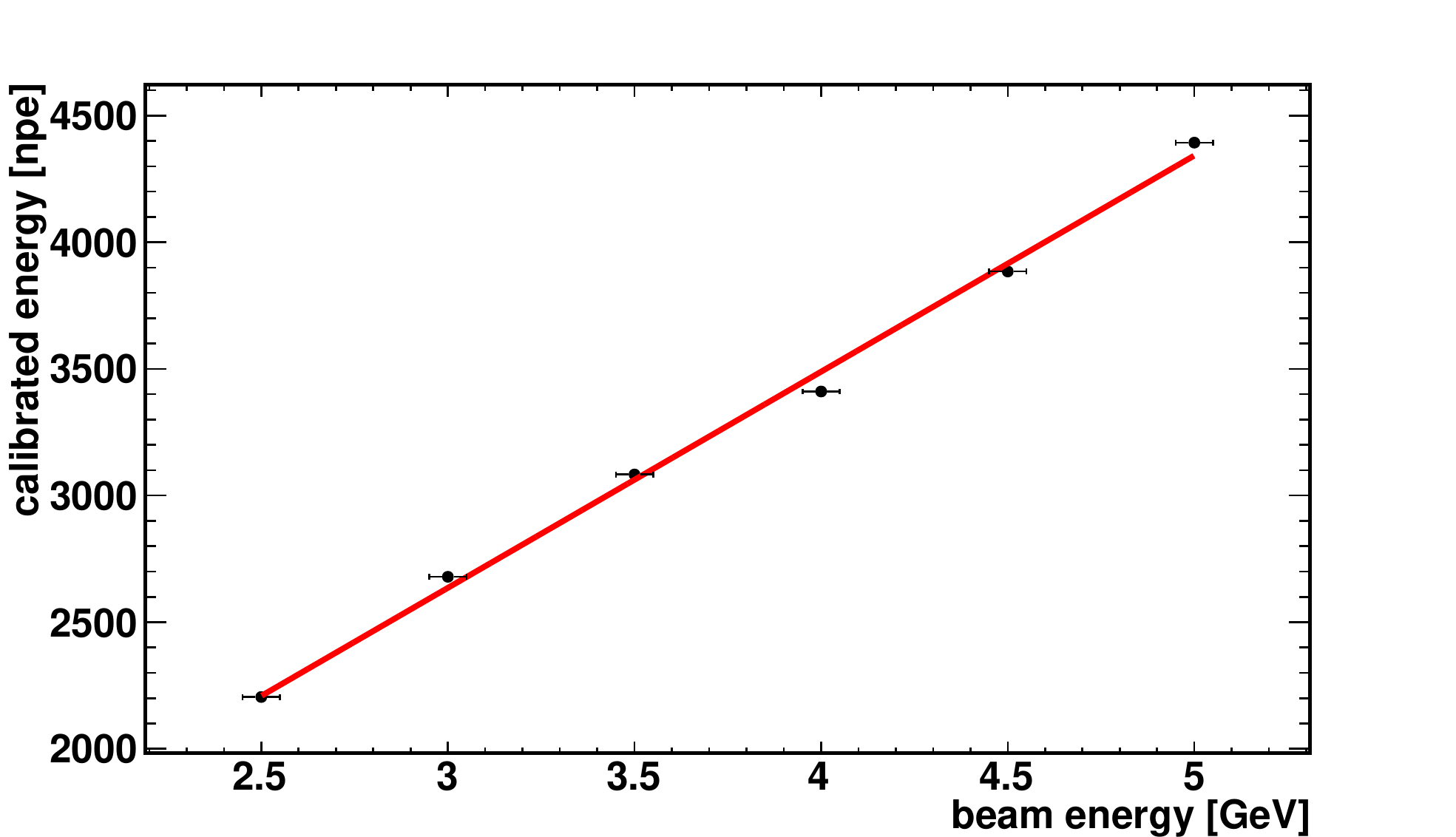}
\caption{\label{fig:calolinear}Reconstructed and calibrated electron energies as a function of nominal beam energy. The uncertainties on the nominal beam energies are set to 50\,MeV; the mean npe uncertainties are negligible. The y-intercept for best-fit line is at $150 \pm 80$\,npe.}
\end{figure}

To extract the energy resolution and linearity functions, beam energies in the range 2.5 to 5\,GeV were used. 
The reconstructed energy distribution for each beam energy was fitted to a Gaussian function to determine the mean and width.
The linearity result is shown in Fig.~\ref{fig:calolinear}. 
The range extends well beyond the decay-energy endpoint for the Muon \gm\ experiment, which is at about 3.1\,GeV.

\begin{figure}[htbp]
\centering
\includegraphics[width=\linewidth, trim=0 0 40 0,clip]{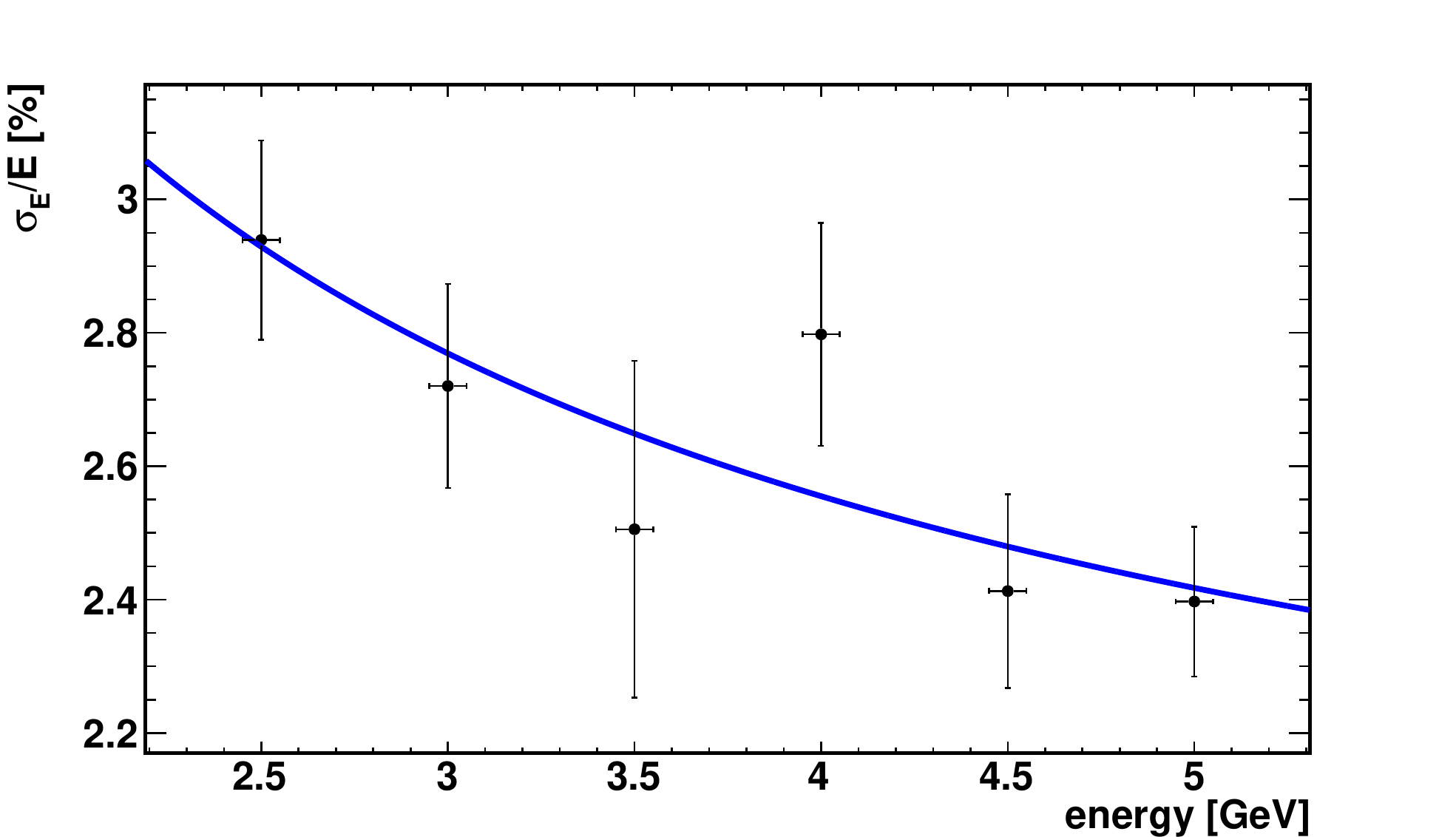}
\caption{\label{fig:caloreso}Energy resolution as a function of beam energy.}
\end{figure}

Figure~\ref{fig:caloreso} shows the dependence of the energy resolution $\sigma_{E}/E$ as a function of electron energy. 
The energy dependence can be parameterized by
\begin{equation}
\frac{\sigma_{E}}{E} = \sqrt{ C_{\rm{E}}^2 + \frac{S_{\rm{E}}^2}{E/\rm{GeV}}}
\end{equation}
where $C_{\rm{E}}$ is a term accounting for the shower containment variance as studied in \cite{Fienberg:2014kka}, and $S_{\rm{E}}$ is a stochastic term describing the statistical fluctuation of the electromagnetic shower. 
The resultant fit is shown in Fig.~\ref{fig:caloreso}. 
The extracted value of $C_{\rm{E}}$ is ($1.86 \pm0.43$)\% and the corresponding value for $S_{\rm{E}}$ is ($3.56 \pm 0.77$)\%.
Based on this parameterization, the energy resolution at 2\,GeV is 3.1\%, surpassing the design target of 5\%. 
While the constant term is consistent with that found in \cite{Fienberg:2014kka}, the resolution is improved by about 11\%.

\subsection{Spatial uniformity}

\begin{figure}[htbp]
\centering
\includegraphics[width=\linewidth]{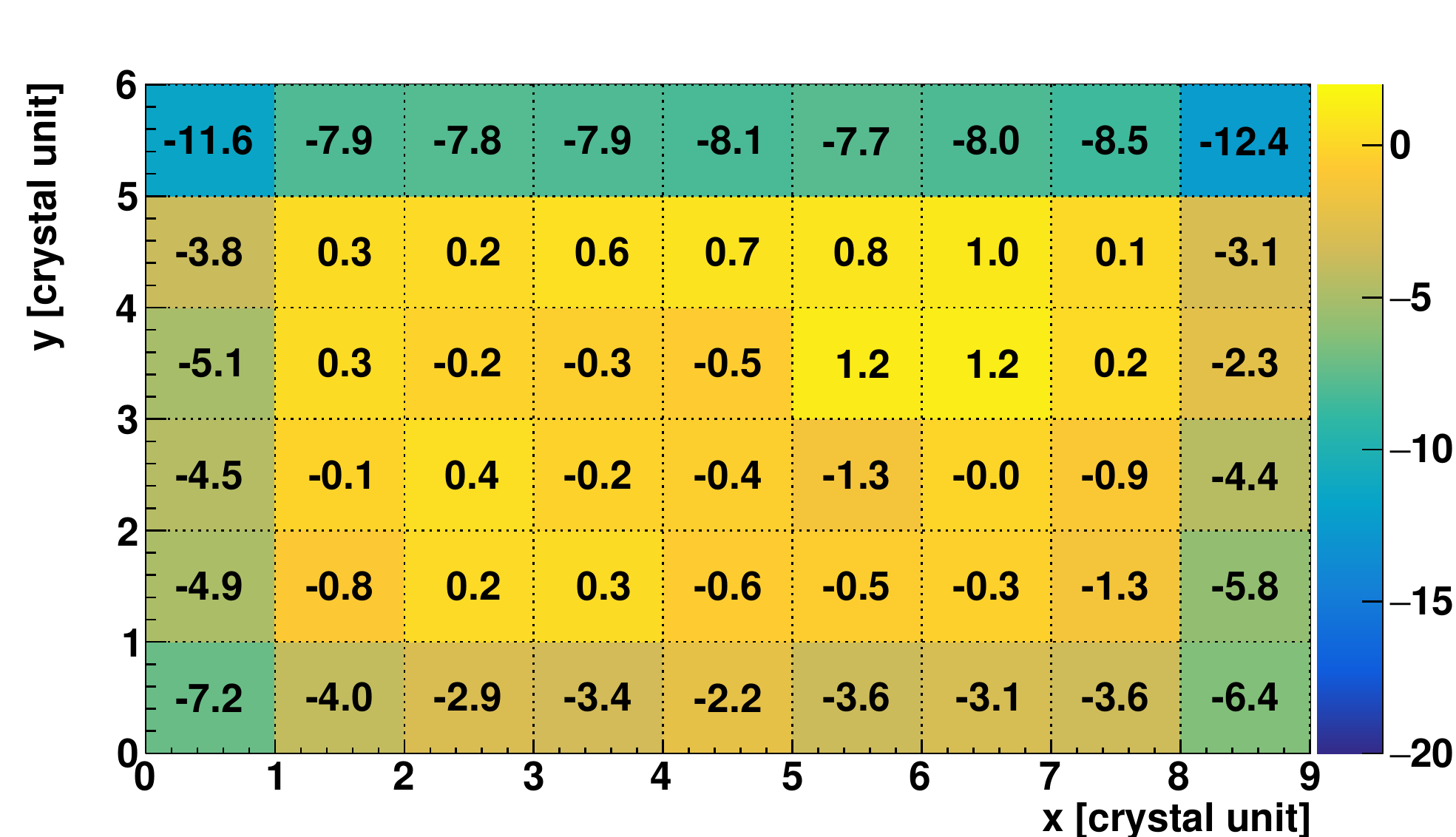}
\caption{\label{fig:calofinescan1}Spatial uniformity of the energy response of a cluster when the electron beam is impinging near the center of each calorimeter channel. The scale is in percent with respect to the average of non-corner channels.}
\end{figure}

\begin{figure}[htbp]
\centering
\includegraphics[width=\linewidth, trim=0 0 40 0,clip]{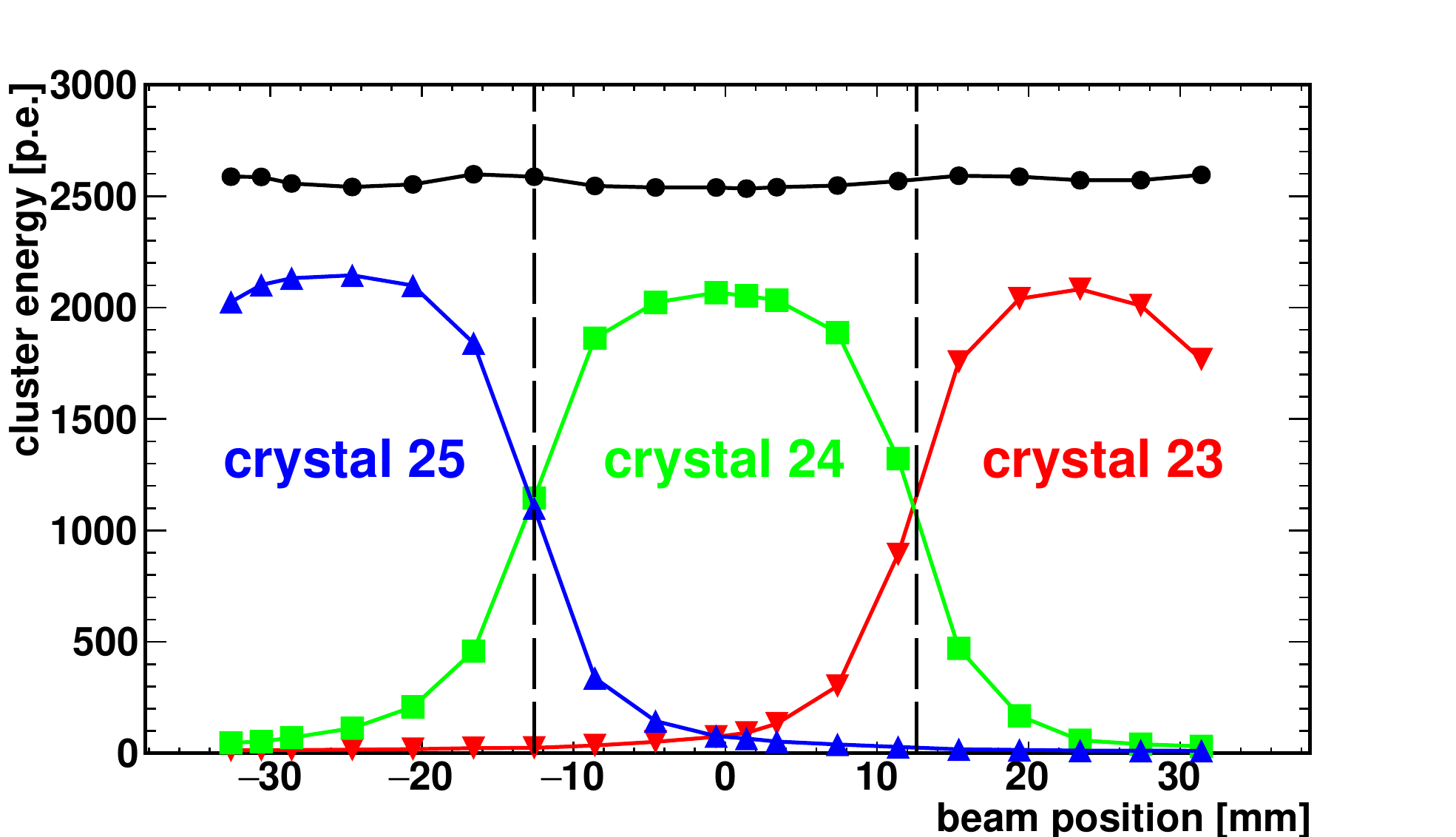}
\caption{\label{fig:calofinescan2}Uniformity of the energy response of the calorimeter tested using horizontal position scan. Full circle is the mean cluster energy, up-pointing triangle (square, down-pointing triangle) is the mean energy registered by crystal 25 (24, 23) as a function beam hit position on the calorimeter. Dashed lines are the border between two crystals.}
\end{figure}

An important aspect of the calorimeter is the spatial uniformity of the response. 
This aspect was studied by comparing the energy response of the calorimeter for various beam impact positions. 
Figure~\ref{fig:calofinescan1} shows the deviation from the mean reconstructed cluster energy in percent for each channel when the electron beam is impinging on the center of each channel. 
The homogeneity is better than 2\% for non-corner channels.
Lower reconstructed energies for corner channels, as well as the uppermost and lowermost rows and leftmost and rightmost columns, are expected due to the leakage of electromagnetic showers.
A finer scan in position is also performed by moving the beam across three channels with a step size of 2-4\,mm as shown in Fig.~\ref{fig:calofinescan2}. 
The reconstructed cluster energy as a function of beam position is stable at the 2\% level.

\section{Summary}
We report on a performance study of the instrumentation for measuring the anomalous precession frequency in the Muon \gm\ experiment at Fermilab.
Unique features of this performance study include fast Cherenkov crystals of \pb; ``PMT-like" pulse width; exclusive use of custom 800\,MSPS waveform digitizers to digitize the SiPM response; the use of hybrid CPU-GPU and MIDAS DAQ system; the use of {\it art}-based event reconstruction framework; and a laser calibration system with a high degree of pulse-to-pulse intensity stability.
Through a series of test beams at SLAC and a final validation presented in this paper, we froze the design of the calorimeter system and went into production for all subsystems. The ``in-ring" performance of the 24-time-larger calorimeter system will be reported in a future publication.

\section*{Acknowledgments}

We warmly thank the whole SLAC ESTB staff, especially Carsten Hast and Keith Jobe
for hosting this effort, which was supported under Department of Energy (DOE) contract DE-AC02-76SF00515. 
We acknowledge the tremendous role of the CENPA and Cornell design and fabrication teams.
We thank Leah Welty-Rieger (Fermilab, USA) for providing Figure 1.
This research was supported by the National Science Foundation (NSF) MRI program (PHY-1337542), by the DOE Offices of Nuclear (DE-FG02-97ER41020) and High-Energy Physics (DE-SC0008037), by the NSF Physics Division (PHY-1205792, PHY-1307328, PHY-1307196, DGE-1144153), by the Istituto Nazionale di Fisica Nucleare (Italy), and the EU Horizon 2020 Research and Innovation Program under the Marie Sklodowska-Curie Grant Agreement No.690385 and No.734303, and by the National Natural Science Foundation of China (11375115) and the Shanghai Pujiang Program (13PJ1404200).

\section*{References}
\bibliography{sections/nim_slac2016}

\end{document}